\documentclass[aps,pra,superscriptaddress,twocolumn,floatfix]{revtex4-1}
\usepackage{amsmath}
\usepackage{float}
\usepackage{graphicx}
\usepackage[T1]{fontenc}
\usepackage{ulem}
\usepackage{braket}
\usepackage{bbold}
\usepackage{hyperref} % demanded by SciPost
\newcommand{\ben}{\begin{enumerate}}
\newcommand{\een}{\end{enumerate}}
\newcommand{\beq}{\begin{equation}}
\newcommand{\eeq}{\end{equation}}
\newcommand{\beqs}{\begin{equation*}}
\newcommand{\eeqs}{\end{equation*}}
\newcommand{\bali}{\begin{align}}
\newcommand{\eali}{\end{align}}
\newcommand{\balis}{\begin{align*}}
\newcommand{\ealis}{\end{align*}}
\newcommand{\beqa}{\begin{eqnarray}}
\newcommand{\eeqa}{\end{eqnarray}}
\newcommand{\beqas}{\begin{eqnarray*}}
\newcommand{\eeqas}{\end{eqnarray*}}
\def\bals#1\eals{\begin{align*}#1\end{align*}}
\def\bal#1\eal{\begin{align}#1\end{align}}
\newcommand{\pe}{\phantom{e}}

\begin{document}

\title{Local orbital formulation of the Floquet theory of projectile 
electronic stopping}

\author{Marjan Famili}
\affiliation{Theory of Condensed Matter, Cavendish Laboratory, 
University of Cambridge, J. J. Thomson avenue, Cambridge CB3 0HE, 
United Kingdom}

\author{Nicol\`o Forcellini}
\affiliation{Theory of Condensed Matter, Cavendish Laboratory, 
University of Cambridge, J. J. Thomson avenue, Cambridge CB3 0HE, 
United Kingdom}
\affiliation{Beijing Academy of Quantum Information Sciences, Beijing 100193, China}
             
\author{Emilio Artacho}
\affiliation{Theory of Condensed Matter, Cavendish Laboratory, 
University of Cambridge, J. J. Thomson avenue, Cambridge CB3 0HE, 
United Kingdom}
\affiliation{CIC Nanogune BRTA and DIPC, Tolosa Hiribidea 76, 
20018 San Sebasti\'an, Spain}
\affiliation{Ikerbasque, Basque Foundation for Science, 
48011 Bilbao, Spain}

\date{\today}

\begin{abstract}

  A recently proposed theoretical framework for the description
of electronic quantum friction for constant-velocity nuclear 
projectiles traversing periodic crystals is here implemented
using a local basis representation.
  The theory requires a change of reference frame to the projectile's,
and a basis set transformation for the target basis functions 
to a ``gliding basis'' is presented, which is time-periodic but
does not displace in space with respect to the projectile,
 allowing a local-basis Floquet impurity-scattering
formalism to be used.
  It is illustrated for a one-dimensional single-band tight-binding 
model, as the simplest paradigmatic example, displaying the 
qualitative behaviour of the formalism.
  The time-dependent non-orthogonality of the gliding basis
requires care in the proper (simplest) definition of a local
projectile perturbation.
  The Fermi level is tilted with a slope given by the projectile
velocity, which complicates integration over occupied states.
  It is solved by a recurrent application of the Lippmann-Schwinger
equation, in analogy with previous non-equilibrium treatment
of electron ballistic transport.
  Aiming towards a first-principles mean-field-like
implementation, the final result is the time-periodic particle density
in the region around the projectile, describing the stroboscopically
stationary perturbation cloud around the projectile, out
of which other quantities can be obtained, such as the
electronic stopping power.
\end{abstract}

\pacs{PACS: }

\maketitle

%%%%%%%%%%%%%%    INTRODUCTION   %%%%%%%%%%%%%%%

\section{Introduction}
  The study of energetic nuclei as projectiles 
shooting through matter has been of great interest for 
over a century \cite{Sigmund,Sigmund2014}. 
  An understanding of the emergent stopping phenomena 
(as the charged particles slow down in matter) from such 
processes is of significant applied interest in a variety of 
contexts, such as  nuclear \cite{nuclear},  
aerospace \cite{aerospace} and medical \cite{medical}.
  It is also of fundamental interest, as 
a canonical problem of quantum systems strongly out of 
equilibrium.

  Electronic stopping processes have been simulated over 
the years using various theoretical frameworks and
approximations.
  From the theoretical side, there are two important 
paradigms for describing electronic stopping in the 
non-relativistic limit. 
  Lindhard's linear response theory \cite{lindhard1954,lindhard1963} 
is applicable to any host material and is accessible 
to first-principles theory \cite{Reining2016}.
  However, it assumes weak effective interaction between 
the projectile and the target electrons, which is a very 
limiting approximation, especially at low velocities 
\cite{race2010}. 
  A fully nonlinear theory was proposed for the
homogeneous electron liquid, including first-principles 
calculations, by Echenique, Nieminen, and Ritchie for the 
low projectile-velocity $v\rightarrow 0$ limit \cite{echenique81}.
  It was later extended to finite $v$ 
\cite{Schonhammer1988,Bonig1989,Zaremba1995,Lifschitz1998},
and it was also generalized to any (non-homogeneous) metal
- still for the low-$v$ limit \cite{Nazarov2005}.
  Both the linear-response and jellium paradigms for
electronic stopping assume a constant-velocity projectile.
  It is a very extended approximation in the community
given the fact that the large projectile mass (as compared
with the electronic) results in a reduction of velocity which is barely 
appreciable in the nano-scale.

  Explicit simulations of the electronic stopping 
processes using time-dependent tight-binding \cite{race2010} 
and time-dependent density-functional theory (TDDFT) are the
state-of-the-art techniques for the treatment of
nonlinear stopping in materials beyond simple metals 
\cite{Pruneda2007,Krasheninnikov2007,Quijada2007, 
Hatcher2008, Correa2012, Zeb2012, Zeb2013, Ojanpera2014, 
Ullah2015,Li2015, Wang2015, Schleife2015, Lim2016,
Quashie2016, Reeves2016, Li2017, Yost2017, Bi2017, 
Ullah2018}.
  However, these calculations remain computationally 
expensive, since the projectile propagates across
a large simulation box containing as much host material
as possible, in periodic boundary conditions.
  In addition to guaranteeing convergence with system
size (minimizing the effect of the multiple replicas 
of the projectile), these simulations rely on the
heuristic ascertaining on having reached
a stationary state. 

  A recent work introduced a theoretical framework which 
allows going beyond both the linear-response and jellium 
approximations in the direct characterisation of the 
stationary state for the study of electronic stopping 
processes \cite{Forcellini2019}.
  It is based on exploiting a discrete translational 
invariance in space-time for ion projectiles moving 
at constant velocity along periodic trajectories in 
crystals.
  When changing reference frame to the one moving with 
the projectile, the problem becomes time periodic and
the theory can be formulated using Floquet theorem 
\cite{Shirley1965,Hanggi1997}.
  It becomes a time-periodic generalisation of
the time-independent problem faced when doing the
same change of reference frame in jellium \cite{echenique81},
now allowing for any periodic potential, and therefore
any crystalline solid of whatever character and chemistry, 
no longer limited to ideal metals.
  The conservation of single-particle (Kohn-Sham particle) 
energy in the scattering processes (in the projectile frame)
in the jellium case now becomes Floquet quasi-energy 
conservation \cite{Forcellini2019}.

  A natural route towards a first principles implementation
of the Floquet theory of electronic stopping is using 
local functions as basis, for reasons analogous to
those that gave very successful local-function implementations
of electronic ballistic transport in the nanoscale
\cite{transiesta, smeagol, gollum}, using 
scattering theory by the means of  Green's functions and 
Dyson's equation.
  Here we propose the main conceptual ingredients for such
an implementation of Floquet stopping theory, setting up the 
paradigm in terms of the simplest possible model: 
a one-band, one-dimensional (1D) tight-binding model,
with a local perturbation moving at a constant velocity
along the system, as established in 
Section~\ref{sec:tb_glidingbasis}.

  The first difficulty is encountered with the local 
functions of the basis moving past the projectile
(at the origin) at a velocity of $-v$.
  This is addressed by introducing a ``gliding'' basis 
transformation to time-periodic but immobile (in space) functions
(Section~\ref{sec:gliding}).
  The adequate description of a projectile local 
perturbation is presented in Section~\ref{sec:projectile},
and the Floquet scattering problem is then solved 
in a Green's function formalism via the Dyson equation
(Section~\ref{sec:scattering_problem}).
  
  An independent-particle formalism is assumed, 
thinking of a mean-field-like implementation
such as Kohn-Sham TDDFT. 
  Single-particle
occupation in this non-equilibrium setting is 
addressed in Section~\ref{sec:occupation}.
  Finally, the time-periodic perturbed particle
density $n(x,t)$ is obtained for the stationary solution
around the projectile -- actually, stroboscopically 
stationary: invariant when looking at it at times 
separated by the relevant time period.

  The electronic stopping power $S_e$ has been the key property 
in comparison with experiments, and it is also important for 
radiation-damage modelling at different length and time scales. 
  It has been conventionally obtained from single-particle
properties \cite{echenique81,Schonhammer1988,Bonig1989,
Zaremba1995,Lifschitz1998}, namely, the scattering
amplitudes and corresponding energy excitations for the
individual scattering processes.
  Notwithstanding its being a remarkably successful approximation,
obtaining the stopping power in terms of the force on the 
projectile appears as a more suitable definition, amenable
to exact treatment under a wider scope of levels of theory.
  That force can be extracted from the particle density
$n(x,t)$, as appears in Section\ref{sec:density}.
  Scattering amplitudes are also computed in the
appendices.

%%%%%%%%%%%%%%   THEORETICAL FRAMEWORK  %%%%%%%%%%%%%%%

\label{method}

\section{Theoretical Framework}
\label{sec:floquet_theory}

%%%%%%%%%%%%%%% FIGURE 1 %%%%%%%%%%%%%%
\begin{figure}[t]
\centering
\includegraphics[width=0.5\textwidth]{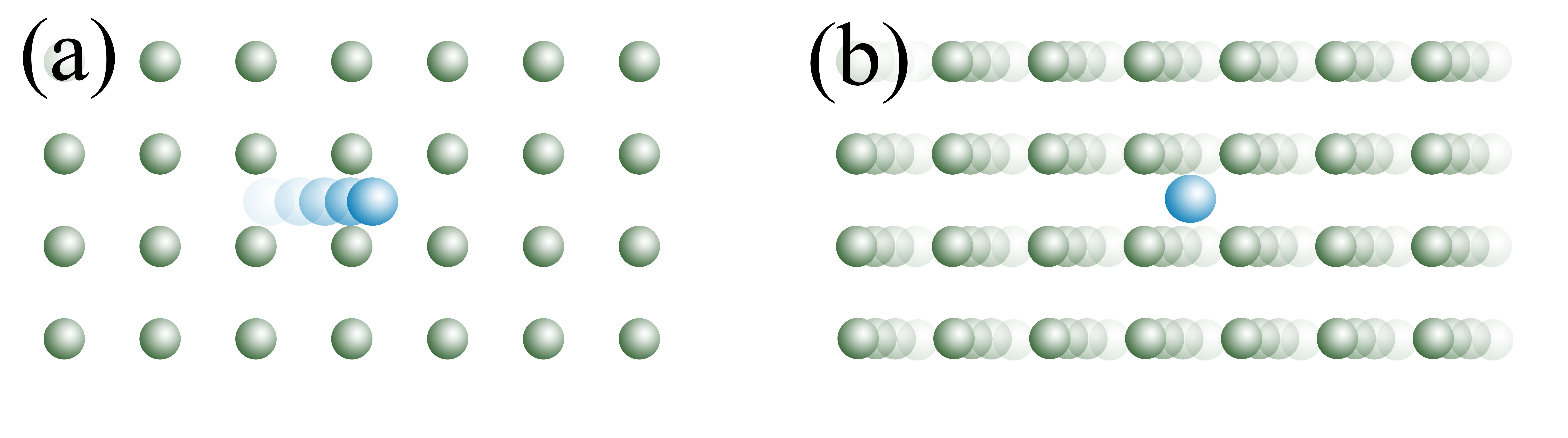}
\caption{Change of reference frame from the laboratory 
reference frame (LRF) (a) to the projectile reference 
frame (PRF) (b) results in a time periodic 
$H(\mathbf{r},t)$ with period $\tau = a/v$, where $a$ is 
the repetition length along the trajectory, and $v$ is
the projectile velocity.}
\label{fig:abstract_fig}
\end{figure}
%%%%%%%%%%%%% END FIGURE 1 %%%%%%%%%%%%

  Consider a projectile moving at constant
velocity $\mathbf{v} = v\hat{\mathbf{v}}$ in the bulk of a 
crystalline solid.
  The constant-velocity projectile is a very extended
assumption in most theoretical approaches to electronic
stopping,  in both linear  and  non-linear response
theories.
  It breaks global energy conservation, the slowing
down of the projectile of the real situation being captured
by the energy uptake of the electrons to a good approximation
for heavy projectiles.
  Following convention we still call it the electronic
stopping problem and associated processes, in spite of
the projectile not slowing down.

  If the motion is along a spatially periodic trajectory of
wavelength $a$, the group of discrete translations 
displacing simultaneously in space by $n a$ and in time 
by $-n v /a$, for any $n\in Z$, leaves the Hamiltonian
invariant.
  This symmetry in space-time can be exploited \cite{Forcellini2019}
through the application of the Galilean transformation 
$\mathcal{G}: \mathbf{r} = \mathbf{r'-v}t$ 
[primed/unprimed indices 
indicating laboratory/projectile frame (LRF/PRF), respectively] 
putting the projectile at rest in the projectile frame.
  The Hamiltonian then takes the form
\begin{equation}\label{eq:periodicham}
H(\mathbf{r},t) = H_0(\mathbf{r}+\mathbf{v}t) + V_P(\mathbf{r}),
\end{equation}
where $H_0$ is the crystal Hamiltonian in the PRF,
and $V_P(\mathbf{r})$ is a local scalar potential representing 
the (now static) projectile. 
  Given the spatial periodicity $a$ along the
projectile trajectory, $H(\mathbf{r},t)$ is time-periodic 
with period $\tau = a/v$.
  Fig.~\ref{fig:abstract_fig} illustrates the boost,
showing how the target atoms move past the projectile.
  In a mean-field setting (such as KS-TDDFT), the projectile
potential itself will also be time dependent, $V_P(\mathbf{r},t)$,
but also time periodic.

  Following \cite{Forcellini2019}, the electronic stopping
problem can be addressed as a time-periodic scattering 
problem for the single-particle %(KS) 
states.
  Floquet's theorem in this context implies
that there are time-dependent solutions of the form
\begin{equation}\label{eq:floquet_solutions}
\Lambda_{n\mathbf{k}}(\mathbf{r},t) = 
e^{-i\varepsilon_n(\mathbf{k}_i)t/\hbar}
\Psi_{n\mathbf{k}}(\mathbf{r},t),
\end{equation}
which represent the stroboscopically stationary 
solutions, where $\Psi_{n\mathbf{k}}(\mathbf{r},t) = 
\Psi_{n\mathbf{k}}(\mathbf{r},t+\tau)$ are
the Floquet \textit{modes}, i.e., the eigenstates 
of the Floquet Hamiltonian $\mathcal{H} = H - 
i\hbar \partial_t$ ($\partial_t \equiv \partial/\partial t$).
  They are labelled by the quasi-momentum $\mathbf{k}$ of
the incoming \textit{unperturbed} Bloch state of the host 
crystal (which becomes Floquet-Bloch in the PRF) with
energy $E_n(\mathbf{k})$ ($n$ being the band index), and 
\begin{equation}\label{eq:quasienergy}
\varepsilon_n(\mathbf{k}) = E_n(\mathbf{k}) -
\hbar\mathbf{k}\cdot\mathbf{v} + mv^2/2
\end{equation}
is the corresponding Floquet quasi-energy 
for a single electron of mass $m$. 

  As usual in scattering theory, the asymptotic form
of the scattering Floquet modes can be expressed
\begin{equation}\label{eq:asymptotic}
\Psi_{n\mathbf{k}}(\mathbf{r},t) \sim \psi_{n\mathbf{k}}(\mathbf{r},t) + 
\sum_{m,\mathbf{k}_f}\mathcal{S}_{n\mathbf{k},m\mathbf{k}_f} 
\psi_{m\mathbf{k}_f}(\mathbf{r},t),
\end{equation}
where $m$, quasi-momenta $\mathbf{k}_f$ and scattering 
amplitudes $\mathcal{S}_{n\mathbf{k},m\mathbf{k}_f}$ 
are determined with scattering theory techniques, including
quasi-energy conservation and outgoing boundary conditions. 
 Throughout the paper, we will use capital letters to distinguish
the perturbed wavefunctions from the corresponding unperturbed 
ones, as in Eq.~\eqref{eq:asymptotic} above, where 
$\psi_{n\mathbf{k}}(\mathbf{r},t+\tau) = \psi_{n\mathbf{k}}
(\mathbf{r},t)$ is the unperturbed Floquet-Bloch mode.
  For more details on the scattering theory for the Floquet-Bloch 
states we refer the reader to \cite{Forcellini2019,ForcelliniThesis}.

\section{Model}

\label{sec:tb_glidingbasis}

  The practical implementation of the above theory based on
a local basis set is tried out in the following.
  A simple tight-binding (TB) model in one dimension (1D) 
serves the purpose of presenting the key concepts and formalisation
needed, and it serves as a paradigmatic example of the 
qualitative physics of the problem.
  In particular, once the reference frame is changed to 
the projectile's, the local basis functions for the target, 
which are static in the LRF, are neither static nor time-periodic,
but displace with velocity $-\mathbf{v}$, and as such, are not
suitable for solving the Floquet scattering problem.
  To address this issue, we propose a basis set
transformation to a set of time-periodic basis states 
(with the same period $\tau$) in Section \ref{sec:gliding}
for the 1D model, which is introduced here first
(for the generalisation to 3D see Appendix \ref{app:tb_general}).

\subsection{One-band moving tight-binding model}\label{sec:1Dtb_intro}

  In the laboratory frame, with one  atom per unit
cell and one orbital per atom, the Hilbert space ${\Omega}'$ 
is spanned by the orthonormal basis set given by the functions 
\beqs
\phi'_{\mu}(x')= \phi'(x'-R'_{\mu})=\langle x' | \phi'_{\mu} \rangle 
\, , \; \; \mu \in \cal{Z}, 
\eeqs   
i.e. atomic orbitals with shape $\phi'(x')$, centered at the
lattice vectors $R'_{\mu} = \mu a$. 
  Prime indices indicate objects in LRF as stated in
Section~\ref{sec:floquet_theory}.
  Assuming only nearest-neighbour hopping of electrons ($\gamma$)
between lattice sites and on-site energy of $\varepsilon_0$,
the Hamiltonian can be written as
\beq\label{eq:h0prim}
H'_0 = \varepsilon_0 \sum_{\mu} \ket{\phi'_{\mu}}\bra{\phi'_{\mu}} - 
\gamma \sum_{\mu}\left( \ket{\phi'_{\mu}}\bra{\phi'_{\mu+1}} 
+ h.c. \right),
\eeq
with $h.c.$ indicating 
the Hermitian conjugate. 
  The eigenvalues and eigenstates of this time-independent 
Hamiltonian satisfying $H'_0\ket{\psi'_k} = E(k)\ket{\psi'_k}$ 
are
\bal
E(k) &= \varepsilon_0-2\gamma \cos{(ka)} \\
\ket{\lambda'_k} &= \frac{1}{\sqrt{N}} \sum_{\mu} e^{ika\mu}
|\phi'_{\mu}\rangle \, ,
\eal
labelled by the crystal momentum $k$, conserved in the 
unperturbed model. $N$ is the number of unite cells in
periodic boundary conditions.
  The quantum number $k$ is not primed, since it 
unequivocally labels the Bloch states in both LRF and PRF. 

  The Bloch waves in the real-space representation and with
explicit time dependence in the energy phase,
\beq
\lambda'_k(x',t) = e^{-iE(k)t/\hbar} \frac{1}{\sqrt{N}} 
\sum_{\mu} e^{ika\mu} \phi'_{\mu}(x') \; ,
\eeq
can be transformed to the PRF via $\mathcal{G}$ 
as \cite{landau}
\beq
\label{eq:movingBloch}
\lambda_k(x,t) = e^{-i \frac{mv}{\hbar}x} e^{-i [E(k) + 
\frac{1}{2} mv^2] t/\hbar}
\frac{1}{\sqrt{N}} \sum_{\mu} e^{ika\mu} \tilde{\phi}_{\mu}(x,t) ,
\eeq
where 
the moving basis functions in PRF are defined as
\beq
\label{eq:movingbasis}
\langle x |\tilde{\phi}_{\mu}(t)\rangle = \tilde{\phi}_{\mu}(x,t) 
\equiv \phi'_{\mu}(x+vt) \; .
\eeq
  Note that in this frame the lattice, the crystalline potential, 
the basis functions and the electrons described by Bloch functions 
are all displacing with velocity $-v$. 
  The Bloch waves transformed through $\mathcal{G}$ have the Floquet 
form $\lambda_k(x,t) = e^{-i\varepsilon(k)t/\hbar}\psi_{k}(x,t)$, where 
$\psi_{k}(x,t)$ is the time-periodic Bloch-Floquet mode
with quasi-energy 
\beq
\label{eq:quasietb}
\varepsilon({k}) = E({k}) -\hbar{k}{v} + mv^2/2 \,.
\eeq
  The Bloch-Floquet modes can be immediately expressed by comparing 
Eq.~\eqref{eq:movingBloch} to the Floquet form, obtaining
\beq \label{eq:BlochFloquetModes}
\begin{split}
\psi_k(x,t) &= e^{-ikvt}e^{-imvx/\hbar}\frac{1}{\sqrt{N}}\sum_{\mu} 
e^{i\mu ka}\tilde{\phi}_{\mu}(x,t)  \\
&= e^{-ikvt}\frac{1}{\sqrt{N}}\sum_{\mu} e^{i\mu ka}  
\phi_{\mu}(x,t) \; .
\end{split}
\eeq
  In the above expression, the phase $e^{imvx/\hbar}$
was absorbed into the local basis $\phi(x,t)$, 
defining the new basis set as the set defined by
\beq
\label{eq:basis-notilde}
\phi_{\mu}(x,t) = e^{-imvx/\hbar}\tilde{\phi}_{\mu}(x,t)
\quad \forall \mu \in \mathcal{Z} \,. 
\eeq 

  The time-periodic function of Eq.~\eqref{eq:BlochFloquetModes} 
defines a Floquet mode, which is an eigenstate of the 
Floquet Hamiltonian $\mathcal{H}(x,t) = H(x,t)
-i\hbar{\partial_t}$ with eigenvalues $\varepsilon(k)$ of
Eq.~\eqref{eq:quasietb}, where $H_0(x,t)$ is
the real-space representation of the TB Hamiltonian 
of Eq.~\eqref{eq:h0prim}, transformed into the moving frame.
  This is true by construction, but can also
be explicitly verified (see Appendix~\ref{app:movsol}). 
  This simple result is key to the solution of the 
Bloch-Floquet scattering problem: By knowing the unperturbed
Bloch-Floquet modes, the allowed asymptotic states of the 
single-particle Bloch-Floquet states are known from
the start, since they have to satisfy quasi-energy 
conservation \cite{Forcellini2019, ForcelliniThesis}.

\subsection{Projectile potential}\label{sec:projectile}

  The system to be studied is that of a constant velocity 
projectile moving along the 1D crystal, 
  A simple tight-binding representation of such a potential 
in the LRF would be
\beq
\label{eq:projectile-LRF}
V_P = \left \{ 
\begin{array}{cl}
|\phi'_{\mu}\rangle \varepsilon_p \langle \phi'_{\mu}|  
& \; ,  \;  t \in [\mu \tau,(\mu+1)\tau) \\ \\
0  & \; , \; t \notin [\mu\tau,(\mu+1)\tau) \; ,
\end{array} 
\right .
\eeq
which represents a constant on-site shift by $\varepsilon_p$ 
on the site the projectile is on, for the duration of its 
passage, i.e. the period $\tau=a/v$, after which it shifts 
to the adjacent site on the right (left) if the projectile 
velocity $v$ is positive (negative).
  An alternative procedure to define the projectile potential 
operator directly on the projectile reference frame is discussed 
below (in Section~\ref{sec:glide-projectile}).

\section{Gliding basis}
\label{sec:gliding}

%%%%%%%%%%%%%%% FIGURE 2 %%%%%%%%%%%%%%
\begin{figure}[t] % it can be [tbxH!] 
\includegraphics[width=0.45\textwidth]{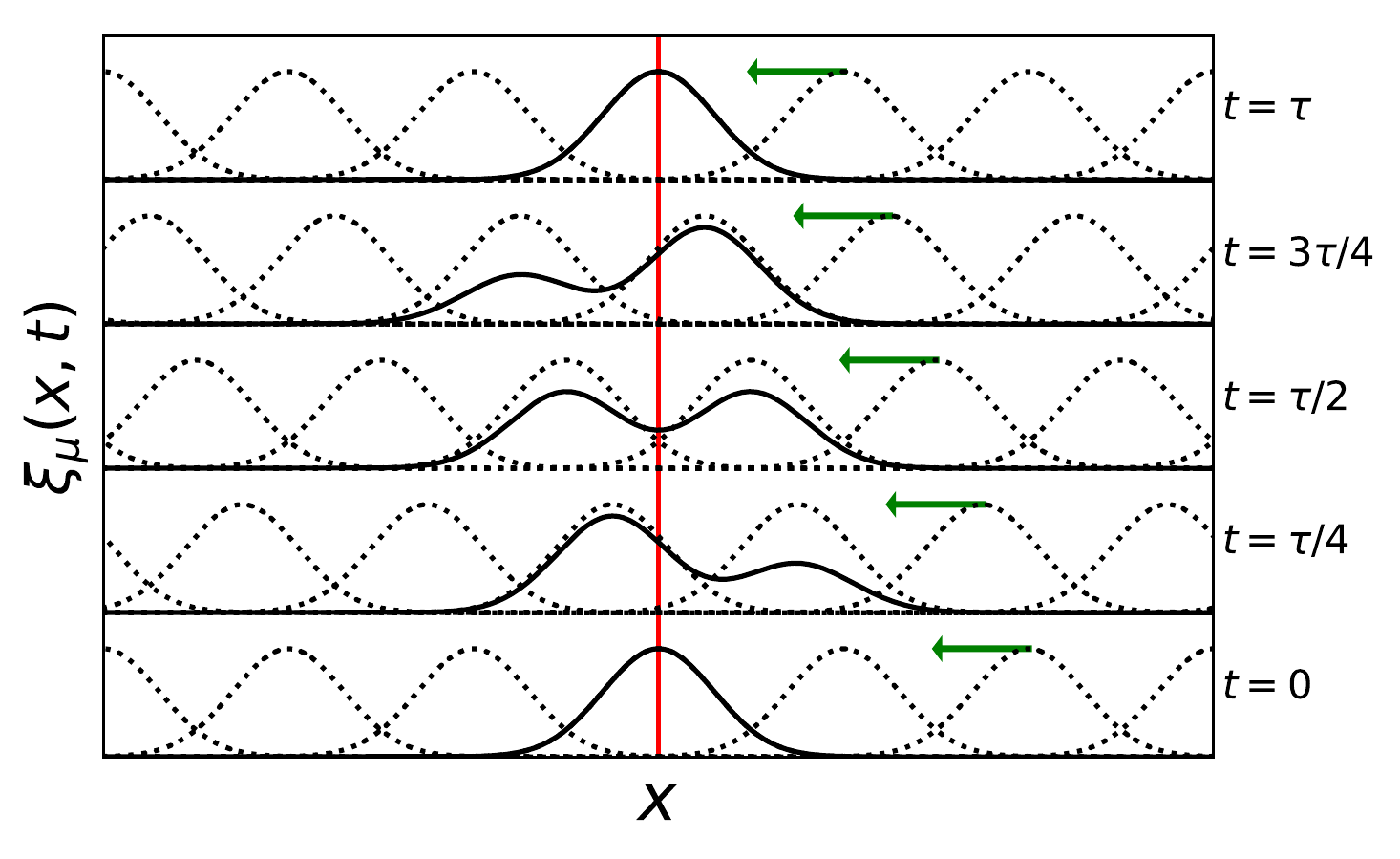}
\caption{Moving original tight-binding basis orbitals 
(dotted lines), and gliding basis function at site $\mu = 0$, 
for four time snapshots (continuous line)
within a period. The red line 
indicates $x=0$, % which is defined to be 
the centre of the projectile in the PRF.}
\label{fig:GlidingBasis}
\end{figure} 
%%%%%%%%%%% END FIGURE 2 %%%%%%%%%%%%%%

  We have been able to state the Floquet modes
of the unperturbed system in terms of the original 
tight-binding basis even though
the basis functions are not periodic themselves.
  This was because we were using the Bloch functions
directly, which are quite close to the Floquet modes.
  The scattering formalism used, will need a local 
basis when dealing with the Floquet modes for the 
total Hamiltonian and the local perturbation induced by 
the projectile.
  The fact that the individual basis functions pass
by the projectile once, never to return, makes them
quite inconvenient.

  A straightforward solution to that problem is the
relabelling of the basis functions every period, as
\beq
\label{eq:step-gauge}
\xi_{\mu}(x,t) = \phi_{\mu+n}(x,t) \, ,
\eeq
where we have defined $n$ from $t=n\tau + \delta t$,
and $\delta t = t \mod \tau$.
  It can be also expressed as
\beqs
\xi_{\mu}(x,t) = \phi_{\mu}(x,\delta t) \, .
\eeqs

  The $\xi_{\mu}$ basis functions are time-periodic 
with period $\tau$, as intended, and are localised
in space on the same lattice as the original one, 
but are now statically defined in the PRF.
  However, the time dependence
is markedly discontinuous, with the basis function
continuously moving leftwards (for $v>0$) during a 
period, at the end of which it performs a sudden jump 
rightwards to start again.
  Such behaviour will be hard to converge in the 
Fourier expansions to be performed below.
 
  A transformation to a basis with smoother time 
dependence is proposed here for numerical convenience, 
each basis function gradually morphing onto its neighbour 
on the left (right) for $v>0$ ($v<0$),
so that the label reassignment happens smoothly.
  Such procedure gives rise to the time-periodic,
non-orthogonal gliding basis illustrated in 
Fig.~\ref{fig:GlidingBasis}, which is also defined on the 
static lattice in PRF, and which can be expressed as
\beq
\label{eq:gliding}
\xi_{\mu}(x,t) = {\cal N}(t) 
[ f(\delta t) \, \phi_{\mu + n} (x,t) \! + \!  
f(\delta t \!\!-\!\!\tau) \, \phi_{\mu +n+1} (x,t)] 
\eeq
for $\delta t \in [0,\tau]$, and $t=\delta t + n \tau$, 
and with
\beqs
{\cal N}(t)= \left [ \; |f(\delta t)|^2 + 
|f(\delta t\!\!-\!\!\tau)|^2 \; \right ]^{-1/2}  \; 
\eeqs
defined as the normalisation at all times.
  It can also be written as
\beq
\label{eq:gliding-alt}
\xi_{\mu}(x,t) = {\cal N}(t) 
[ f(\delta t) \, \phi_{\mu} (x,\delta t) \! + \!  
f(\delta t \!\!-\!\!\tau) \, \phi_{\mu+1} (x,\delta t)] \, .
\eeq

  The function $f(t)$ which defines the basis transformation, 
should be non-zero only in the $[-\tau,\tau)$ interval, 
\beqs
f(t) = \left \{ 
\begin{array}{ll}
 \tilde{f}(t)  & \; ,  \;  t \in [-\tau,\tau) \\
 0  & \; , \; t \notin [-\tau,\tau)
\end{array} 
\right .
\eeqs
  Although it is not necessary, it is numerically 
convenient to ensure continuity (and hopefully smoothness)
of the function at $t=\pm\tau$. 
  Fig.~\ref{fig:GlidingBasis} illustrates the evolution of such 
a basis function.
  Note the use of $\phi(x,t)$ from Eq.~\eqref{eq:basis-notilde}
in this definition.   

  Since the basis given by the set $\{ \xi_{\mu}(x,t), \forall \mu 
\in \mathcal{Z}\}$ spans the same space as spanned by 
$\{ \phi_{\mu}(x,t), \forall \mu \in \mathcal{Z}\}$, 
the shape of $f(t)$ represents a gauge freedom, which can be 
exploited for practical considerations such as maximising 
smoothness for Fourier transform truncation or simplicity 
in the equations.
  Examples of $\tilde{f}(t)$ can be found in Appendix~\ref{app:fs}.
  The numerical calculations in this paper are done using 
the gauge function $\tilde{f}_2(t)$ in Eq.~\eqref{eq:gauge}, which 
ensures the continuity of the gauge function as well as 
of its first derivative, while the time-discontinuous 
transformation giving the simplest formalism (of 
Eq.~\eqref{eq:step-gauge}) is given by the gauge step 
function of $\tilde{f}_4(t)$ in Eq.~\eqref{eq:gauge}.

\subsection{Overlap and Hamiltonian in gliding basis}

  The gliding basis set $\{|\xi_{\mu}\rangle \}$ is non-orthogonal.
  Its overlap matrix, or metric tensor, $S_{\mu\nu} = \langle 
\xi_{\mu} (t) | \xi_{\nu} (t) \rangle$ is given by
\begin{equation}\label{eq:overlap1}
S_{\mu\nu}(t) = \delta_{\mu\nu} + s(t) \,\,
(\delta_{\mu,\nu+1} + \delta_{\mu,\nu-1}) \; ,
\end{equation}
where we assumed that $f(t)$ is real, and defined $s(t)$ as
\beqs
s(t) = {\cal N}(t)^2 \,\, f(\delta t) f(\delta t-\tau) \; .
\eeqs

  The unperturbed Hamiltonian in Eq.~\eqref{eq:h0prim} can be 
expressed in the gliding basis when transferred to PRF.  
  $H_0(t)$ is a tridiagonal matrix, with time-periodic 
sub- and supra-diagonals, which  annihilate after each period. 
  The non-zero matrix elements are
\beqs
\langle \xi_{\mu}, t | H_0 | \xi_{\mu}, t \rangle = \varepsilon_0 % \,
-2 \gamma s(t)
\eeqs
\beqs
\langle \xi_{\mu}, t | H_0 | \xi_{\mu +1}, t \rangle = % & = 
\langle \xi_{\mu}, t | H_0 | \xi_{\mu-1}, t \rangle  % \\
= \varepsilon_0 s(t) - \gamma \, ,
\eeqs
and
\beqs
\langle \xi_{\mu}, t | H_0 | \xi_{\mu +2}, t \rangle = % & =
\langle \xi_{\mu}, t | H_0 | \xi_{\mu-2}, t \rangle % \\
= - \gamma s(t) \, ,
\eeqs
although we will not need to solve for $H_0$ given that we already
have the unperturbed (asymptotic) Floquet scattering modes
from Eq.~\eqref{eq:BlochFloquetModes}.

\label{sec:glide-projectile}
  The projectile potential $V_P$ of Eq.~\eqref{eq:projectile-LRF} 
becomes a matrix with elements
\bal 
\langle \xi_0, t | V_P | \xi_0, t \rangle &= \varepsilon_p \, 
\mathcal{N}(t)^2 |f(\delta t)|^2 
\nonumber \\
\langle \xi_1, t | V_P | \xi_1, t \rangle &= \varepsilon_p \, 
\mathcal{N}(t)^2 |f(\delta t-\tau)|^2 
\nonumber \\
\langle \xi_0, t | V_P | \xi_1, t \rangle &= \varepsilon_p \, 
\mathcal{N}(t)^2 f(\delta t)f(\delta t-\tau) 
\nonumber \\
\langle \xi_1, t | V_P | \xi_0, t \rangle &= \varepsilon_p \, 
\mathcal{N}(t)^2 f(\delta t)f(\delta t-\tau)  
\label{Eq:VP_GB}
\eal
and zero otherwise.

  An alternative way of introducing the projectile potential 
is by parametrizing it directly in the gliding basis, already in PRF.
  It is appealing given its conceptual and implementation simplicity.
  The most straightforward choice would be to define $V_P$ by specifying its
representation in the gliding basis as the matrix
\beq
\label{eq:wrong-projectile}
\langle \xi_{\mu}, t | V_P | \xi_{\nu}, t \rangle =
\varepsilon_p \delta_{\mu\nu} \delta_{\mu 0} \, ,
\eeq
that is, a matrix with a constant on-site term at the zero site as the 
only non-zero term. 
  This choice displays, however, two conceptual disadvantages:
($i$) $V_P$ would then be gauge-dependent; a different choice of $f(t)$
in Eq.~\eqref{eq:gliding} not only affects convergence but also the results. 
($ii$) Transforming back to the original basis, it can be shown that the 
decay length of the potential being represented depends on time, and 
actually diverges at $t=\tau/2$.
  It is shown in Appendix~\ref{app:wrong-projectile}. 
  
  Eq.~\eqref{eq:wrong-projectile} can be expressed in operator 
form as
\beqs
V_P = |\xi^0, t \rangle \varepsilon_p \langle \xi^0, t| \, ,
\eeqs
where we are using the instantaneous dual basis $\{|\xi^{\mu}, t\rangle\}$,
defined, as usual (see e.g. in this context \cite{Artacho2017}), 
as the set of states (at any given time) that satisfy
\beqs
\langle \xi^{\mu} | \xi_{\nu}\rangle = \langle \xi_{\nu} | \xi^{\mu}\rangle
= \delta^{\mu}_{\nu} \, , \quad \forall \, \mu, \nu \in Z \, .
\eeqs
  It allows us to extend the proposal to alternative ones using locality in
the natural and matrix representations 
of $V_P$ \cite{Artacho2017}, namely,
\beqs
V_P = |\xi_0, t \rangle \varepsilon_p \langle \xi^0, t| \, ,
\eeqs
and 
\beq
\label{eq:3projectile}
V_P = |\xi_0, t \rangle \varepsilon_p \langle \xi_0, t| \, ,
\eeq
respectively.
  The gauge-dependence problem remains for any of these choices, but the latter
is not affected by the extreme time-dependence of the range of the potential. 
  It has the matrix form
\beq
\varepsilon_p \left (
\begin{array}{ccc}
s(t)^2 & s(t) & s(t)^2 \\
s(t)   &  1   &  s(t)  \\
s(t)^2 & s(t) & s(t)^2 
\end{array}
\right )
\eeq
for the block for $\mu, \nu = -1, 0, 1$, being zero otherwise.
  For this paper, we choose to stay with the definition of $V_P$
given by Eq.~\eqref{Eq:VP_GB}, given its gauge independence. 
  Appendix~\ref{app:wrong-projectile} shows some results for 
the projectile defined as in Eq.~\eqref{eq:3projectile} for comparison,
using the gauge employed throughout this paper.

\subsection{Floquet space}\label{subsec:floquet_space}

  The space spanned by the moving basis set in Eq.~\eqref{eq:movingbasis},
and equivalently, the one spanned by the gliding basis defined in
Eq.~\eqref{eq:gliding}, gives a Hilbert space at time $t$, 
$\Omega(t)$. 
  As an object (for all times) it represents a curved manifold 
\cite{Artacho2017} that satisfies $\Omega(t+n\tau)=\Omega(t)$,
even though the moving basis is not periodic.
  A Floquet space can be constructed as 
${\cal F}=\Omega\otimes {\cal T}$.  
  As geometrical object it would certainly deserve further 
mathematical attention, but, for the purposes of this work, the 
following suffices.

  Consider any time-periodic function $\Phi$ spanned by the moving 
basis in the sense
\beq
\label{eq:periodicphi}
|\Phi (t)\rangle = \sum_{\mu} \Phi^{\mu}(t) |\xi_{\mu} (t) \rangle \; .
\eeq
  Since both $|\psi (t)\rangle$ and all
the $ |\xi_{\mu} (t) \rangle$'s are periodic, then $\Phi^{\mu}(t)$ 
is periodic, too. 
  These coefficients can therefore be expanded as
\beqs
\Phi^{\mu}(t) = \sum_{m = -\infty}^{\infty}  \Phi^{\mu}_m 
e^{im\omega t}
\eeqs
and we can re-express Eq~\eqref{eq:periodicphi} as
\beq
\label{eq:periodiphi2}
|\Phi (t) \rangle = \sum_{\mu,m} \Phi^{\mu}_m e^{im\omega t}  
|\xi_{\mu} (t) \rangle \; .
\eeq
  This expression shows that the Floquet basis set $\{ | \, \xi_{\mu}, 
m \rangle \rangle, \forall \mu,m \in \mathcal{Z} \}$ defined as
\beq
\label{eq:floquetbasis}
\langle \langle x,t \, | \, \xi_{\mu}, m \rangle \rangle = 
\xi_{\mu} (x,t) \, e^{im\omega t} \; ,
\eeq
constitutes a basis that spans the Floquet 
space ${\cal F}$ corresponding to the original basis. 

  The overlap matrix in these basis 
$S_{\mu\nu,mn} = \braket{\braket{\xi_{\mu},m|\xi_{\nu},n}}$
can be expressed as an inner product in the extended 
$\mathcal{F}$ space 
\begin{equation}
\begin{split}
    S_{\mu\nu,mn} &= \frac{1}{\tau}\int^{\tau} S_{\mu\nu}(t)
    e^{-i(m-n)\omega t} dt \\
    &= \delta_{\mu,\nu}\delta_{m,n} + s_{m-n}
    (\delta_{\mu,\nu+1} + \delta_{\mu,\nu-1}),
    \end{split}
\end{equation}
with $s_{m-n} = \frac{1}{\tau}\int^{\tau} dt \ s(t)
e^{-i(m-n)\omega t} $.
Similarly for $H$,
\begin{equation}
H_{\mu\nu,mn} = \frac{1}{\tau}\int^{\tau} H_{\mu\nu}(t)
    e^{-i(m-n)\omega t} dt \, .
\label{eq:H_mu_nu_mn}
\end{equation}
The matrix elements of the Floquet Hamiltonian
$\mathcal{H} = H - i\hbar \partial_t$ are
\beq
\mathcal{H}_{\mu\nu,mn} = H_{\mu\nu,mn} 
- i\hbar D_{\mu\nu,mn} +
n \, \hbar\omega \, S_{\mu\nu,mn} \, ,
\label{eq:F_mu_nu_mn}
\eeq
where $D_{\mu\nu,mn}=\frac{1}{\tau}\int^{\tau} D_{\mu\nu}(t)
e^{-i(m-n)\omega t} dt$, and $D_{\mu\nu}(t) = \langle 
\xi_{mu}(t)|\partial_t \xi_{\nu}(t)\rangle$ is the connection
in the manifold \cite{Artacho2017}.
 Since the solutions for the unperturbed
Floquet Hamiltonian (containing the $\partial_t$ term) will be directly
obtained from the $\mathcal{G}$ boost of the Bloch solutions of the
crystalline system, as shown in the next section, 
the calculation of the connection will not be needed.

\begin{figure}[t] % it can be [tbxH!] 
\includegraphics[width=0.45\textwidth]{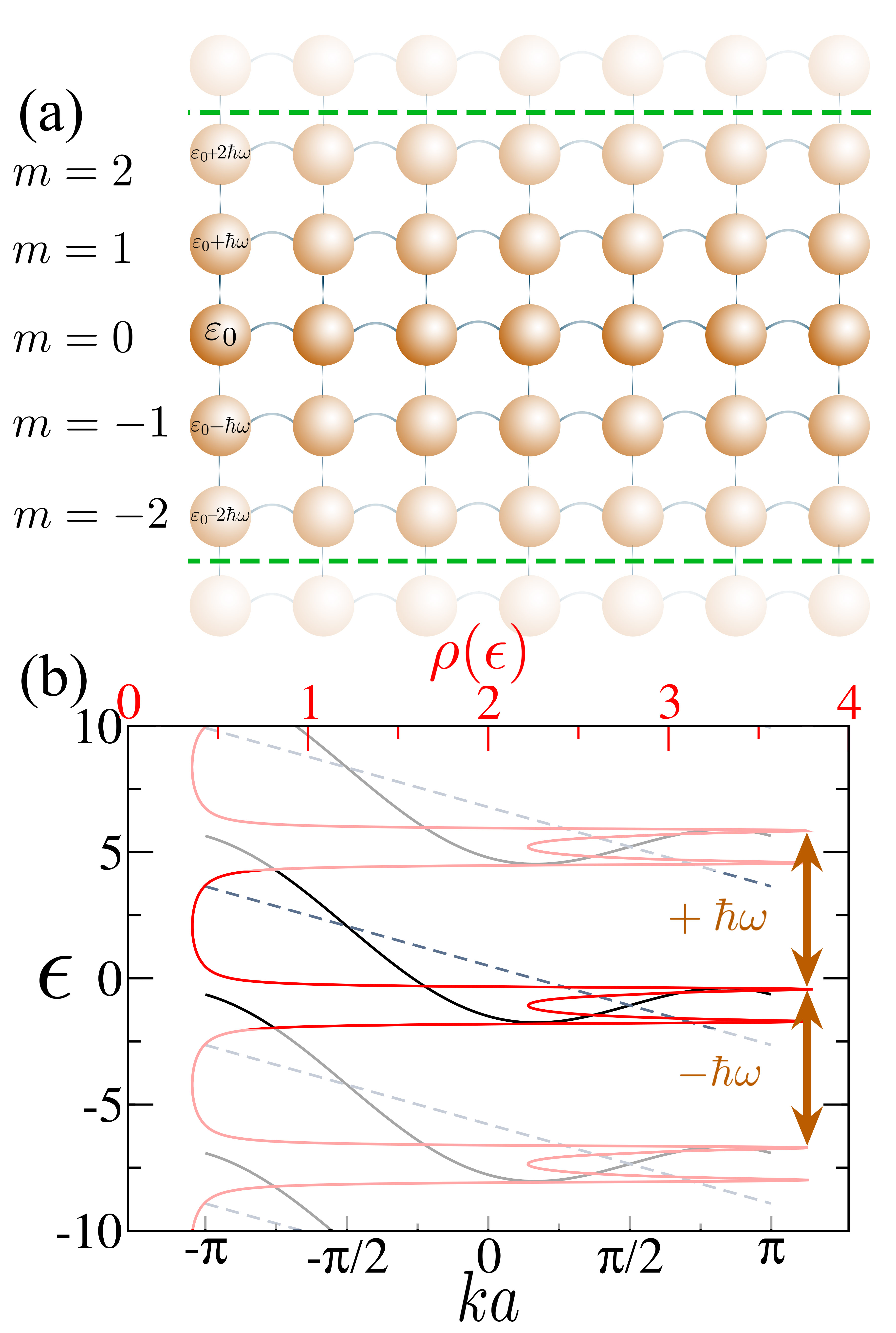}
\caption{ (a) Schematic illustration of a periodically 
driven 1D Floquet chain represented in a 2D lattice.
  The green dashed lines mark the cut-off of
Fourier components to contribute to the primitive cell
in quasi-energy $\varepsilon$ around $\varepsilon=0$.
 (b) Quasi-energy bands vs Bloch momentum $k$ (solid black,
lower abscissa),
and as density of states (red lines, upper abscissa) for the 
unperturbed tight-binding chain moving with 
velocity $-v$ in the projectile reference frame,
for $v=v_0$ in the units given by the hopping energy 
$\gamma$, $\hbar$, and the lattice parameter $a$, 
as $v_0=\gamma a/\hbar$.
  The tilted dashed line represents the Fermi level,
describing half-filling in equilibrium in the LRF.
  The replicas at quasi-energies beyond the chosen cell
are indicated as faded, but shown to illustrate periodicity.}
\label{fig:Fbands}
\end{figure}

  The Hamiltonian of the 1D-chain has been extended in Floquet  
space, Eq.~\eqref{eq:F_mu_nu_mn}, as reflecting a 2D system
[see Fig. \ref{fig:Fbands}(a)].
  The quasi-energy spectrum of Floquet modes is periodic in 
quasi-energy, with an $\omega$ repetition analogous to 
the periodicity in reciprocal space for crystals.
  For the calculation of the modes in a quasi-energy
unit cell (say, around $\varepsilon=0$, as Brillouin zone analog),
the weight on basis functions diminishes with growing $|m|$,
and a cutoff $m_c$ can be established, reducing the 2D 
system to a ribbon of $2m_c$ width, illustrated in 
Fig.~\ref{fig:Fbands}(a).
  We now address the Floquet scattering problem in
the representation given by the Floquet (Fourier) basis
of Eq.~\eqref{eq:floquetbasis}.

%%%%%% SECTION V SCATTERING PROBLEM

%%%%%%%%%%%%%%%%% FIGURE 3 %%%%%%%%%%%%%%%%%%%
%%%%%%%%%%%%%%% END FIG 3 %%%%%%%%%%%%%%%%%%%%

\section{Scattering problem}\label{sec:scattering_problem}

%%%%%%%%%%%%%%%% FIGURE 4 %%%%%%%%%%%%%%%%%%%%
\begin{figure*}[t]
\centering
\includegraphics[width=1\textwidth]{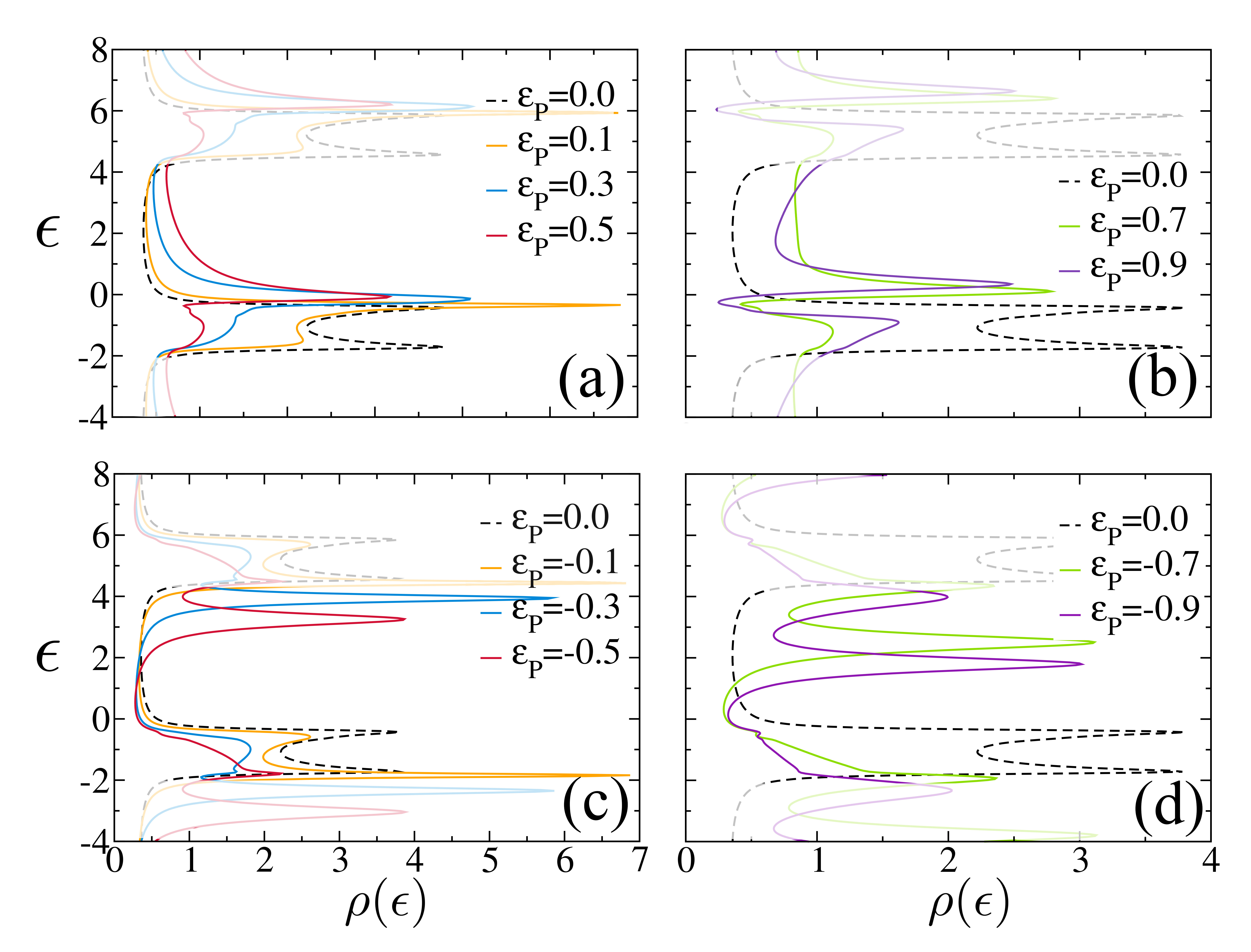}
\caption{Perturbed density of states (DOS) $\rho$ vs quasi-energy
$\epsilon$ on site $\mu=0$ for (a) repulsive weaker 
($\varepsilon_p=0.1,0.3,0.5$,in units of hopping energy $\gamma$) and 
(b) stronger ($\varepsilon_p=0.7,0.9$) projectile potential. 
  Same in (c) and (d) for attractive projectile potential.
The unperturbed DOS is indicated by dashed lines.
  Faded region of the curves are for  the replicas as in 
Fig.~\ref{fig:Fbands}(b).
  The value of $\eta=0.07\gamma$ is used throughout.}
\label{fig:pert_dos_pot}
\end{figure*} 
%%%%%%%%%%%%%%%%% END FIG 4 %%%%%%%%%%%%%%%%%%

\subsection{Asymptotic states in the gliding basis}

  The Bloch-Floquet asymptotic scattering modes of the 
moving tight-binding chain (Eq.~\eqref{eq:BlochFloquetModes}) 
can be expressed in the gliding basis set as defined in
Eq.~\eqref{eq:gliding}, giving
\begin{equation}\label{eq:BlochFloquetModesGliding}
\begin{split} 
\ket{\psi_k(t)} &=   
\frac{1}{\sqrt{N}} \sum_{\mu} \psi^{\mu}_k(t)  
|\xi_{\mu}(t)\rangle \; ,
\end{split}
\end{equation}
where $\psi^{\mu}_k(t)$ are time-periodic coefficients, 
as in Eq.~\eqref{eq:periodicphi}.
  They are phase factors ($|\psi^{\mu}_k(t)|^2=1$)
given by
\begin{equation}\label{eq:alpha_timedependent}
\psi^{\mu}_k(t) = \psi^{\mu}_k(\delta t) = 
\frac{e^{-ik v \delta{t}}e^{ik a \mu}}
{{\cal N}(\delta t) \big[f( \delta t) \, \! 
+ e^{-ik a}f( \delta t-\tau)  \big]} \, .
\end{equation}
  Again, $\delta t = t- n\tau$, with $n$ counting 
the number of periods from $t = 0$.
  The result in Eq.~\eqref{eq:alpha_timedependent}
is to be expected from the fact that the $|\psi_k(t)\rangle$
represent, at any given time, Bloch states of the 
static lattice in PRF, to the points of which each
$|\xi_{\mu}(t)\rangle$ is associated
(the denominator representing the usual normalisation
factor of Bloch states for a non-orthogonal basis).
  The extra phase $e^{-ik v \delta{t}}$
resulting from the transformation goes beyond that
argument. However, it should not be neglected in spite
of its inconvenient discontinuity, which seems
to be an inescapable manifestation of the 
relabelling of basis states at every period.

  Expanding the modes of Eq.~\eqref{eq:BlochFloquetModesGliding}
in the Floquet-Fourier basis $\{ | \, \xi_{\mu}, m \rangle \rangle,
\forall \mu,m \in \mathcal{Z} \}$  yields
\begin{equation}\label{eq:FBexpansion}
\ket{\psi_k}\rangle = \sum_m \psi_{m k}^{\mu} | \, 
\xi_{\mu}, m \rangle \rangle \, ,
\end{equation}
with $\psi_{m k}^{\mu}$ 
defined as
\begin{equation}
\psi_{m k}^{\mu}=\langle \langle \xi^{\mu}, m | 
\psi_k\rangle\rangle = \frac{1}{\tau} \int^\tau  \psi^{\mu}_k(t)   
e^{-im\omega t} dt \, . 
\end{equation}

\subsection{Unperturbed Green's function}

  The scattering problem is addressed here
using the Green's functions $G(\epsilon)$ 
defined for the Floquet Hamiltonian $\mathcal{H}$
as a function of the quasi-energy value $\epsilon$.
  It is analogous to time-independent (energy-conserving)
scattering problems addressed using $G(E)$ where  the energy $E$ is 
the conjugate of time. 

However, in our periodic case, time has 
become as space-like variable for the eigenproblem
being faced for $\mathcal{H}$, the quasi-energy
$\epsilon$ becoming the conjugate of an auxiliary time
$t'$ in the so-called $t,t'$ formalism \cite{Martinez_2003},
which allows for a generalisation of the scattering
formalism involving the Dyson equation in the 
extended Floquet space $\mathcal{F}$.

%%%%%%%%%%%%%%%% FIGURE 5 %%%%%%%%%%%%%%%%%%%%
\begin{figure*}[t]
\centering
\includegraphics[width=1.0\textwidth]{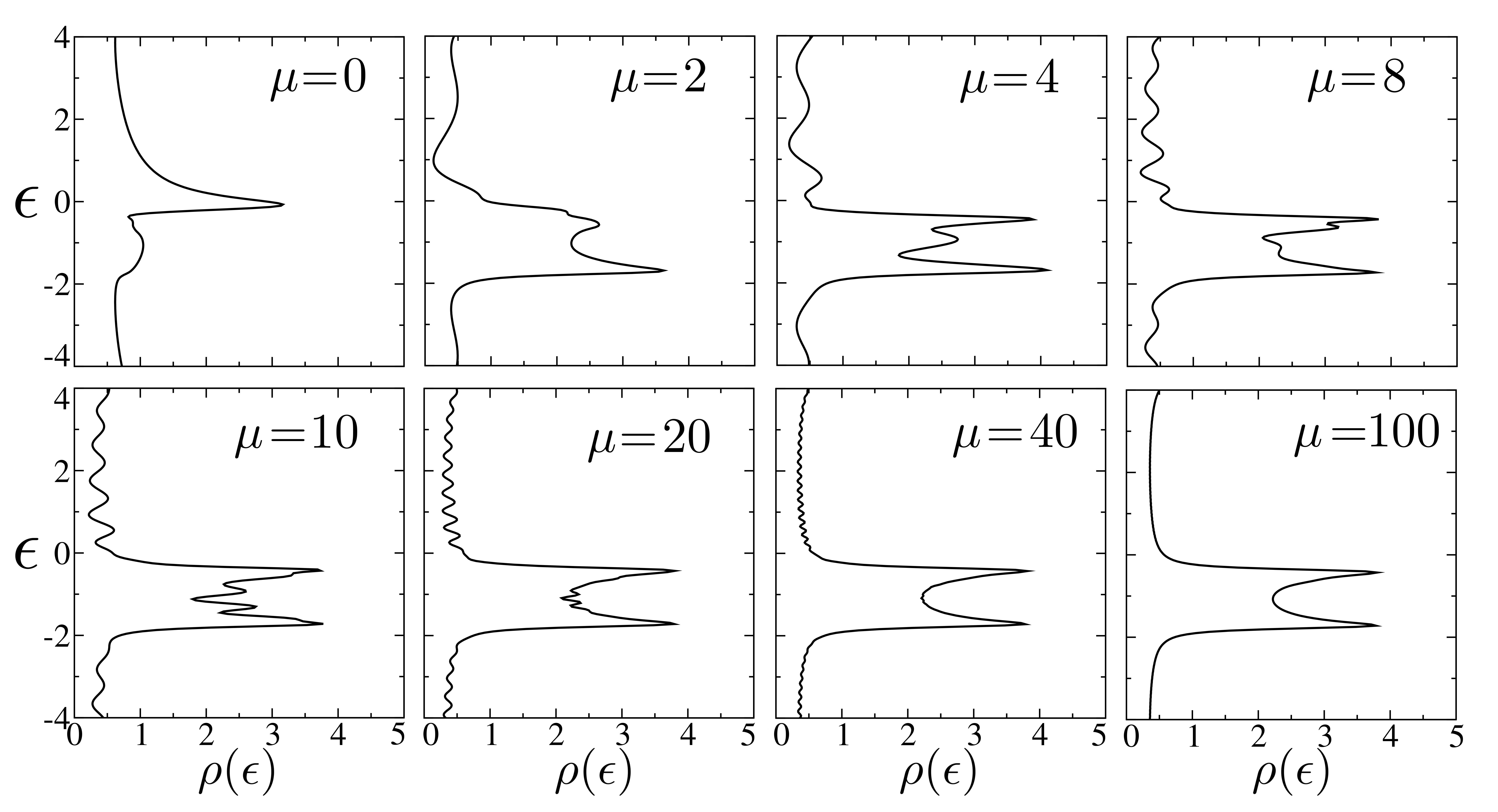}
\caption{Perturbed density of states $\rho$ vs quasi-energy
$\epsilon$ evaluated at various sites $\mu$ moving away from the 
projectile (at $\mu=0$).}
\label{fig:pert_dos_site}
\end{figure*}
%%%%%%%%%%%%%%%%% END FIG 5 %%%%%%%%%%%%%%%%%%

  Knowing the exact eigenstates of $H_0$, and therefore,
the unperturbed Floquet-Bloch states of Eq.~\eqref{eq:FBexpansion}, 
the unperturbed retarded Green's function can be readily written as
a matrix in $\mathcal{F}$ as
\beq  
\label{eq:gmat}
g^{\mu\nu}_{mm'}(\epsilon) = \sum_{k} \frac{\langle\langle
\xi^{\mu}, m | \psi_{k}\rangle\rangle\langle\langle 
\psi_{k} | \xi^{\nu}, m'\rangle\rangle}
{\epsilon -\varepsilon(k)+i\eta} \; ,
\eeq
for $\eta \rightarrow 0^+$, or
\beq  
\label{eq:gmat2}
g^{\mu\nu}_{mm'}(\epsilon) = \frac{1}{N} \sum_{k} \frac{
\psi^{\mu}_{m k} \psi^{\nu *}_{m'k}}
{\epsilon -\varepsilon(k)+i\eta} \; ,
\eeq
where
\beq
\psi^{\mu}_{m k}=\langle\langle
\xi^{\mu}, m | \psi_{k}\rangle\rangle
\eeq
are the expansion coefficients defined
in Eq.~\eqref{eq:FBexpansion}. 

  The unperturbed density of states (DOS) $\rho^0(\epsilon)$ 
of the moving tight-binding chain is then obtained via
\beq
\label{eq:DOS}
\rho^0(\epsilon) = -\frac{1}{\pi} \text{Im} 
\left[\sum_{m m'} \sum_{\mu\nu} g^{\mu\nu}_{m m'} (\epsilon)
S_{\nu\mu,m'm} \right].
\eeq 
  Figure \ref{fig:Fbands} shows $\rho^0(\epsilon)$ in 
the absence of a projectile where the lattice is 
moving with $v=-v_0$, with $v_0=\gamma a/\hbar$.
  The DOS is periodically repeated in $\epsilon$
with a period of $\hbar \omega$, due to the
structure of the quasi-energy spectrum, as 
apparent for the Van Hove singularities 
appearing for Bloch states with a group velocity
equal to the projectile's velocity $v$
(zero velocity in PRF, see Fig.~\ref{fig:Fbands}).

\subsection{Projectile perturbation: Dyson equation} 

  The effect of the projectile is obtained to all orders
using the Dyson equation as for any scattering problem
\cite{Economou}, 
\beq 
\label{eq:dyson_eq0}
\mathbf{G}(\epsilon) = \mathbf{g}(\epsilon) + 
\mathbf{g}(\epsilon) \mathbf{V}_P \mathbf{G}(\epsilon)
\eeq
as expressed as matrices in an abstract form, being 
$\mathbf{G}(\epsilon)$ the perturbed Green's function.
  It can expressed in the computationally convenient way
\beq
\mathbf{G}(\epsilon)=\left[\mathbf{g}^{-1}(\epsilon)- 
\mathbf{V}_P \right]^{-1} \, ,
\label{eq:dyson_eq}
\eeq
which, given the structure of Eq.~\eqref{eq:dyson_eq0},
can actually be solved as a matrix inversion of the matrix 
blocks corresponding to non-zero $\mathbf{V}_P$ elements,
which, from Eq.~\eqref{Eq:VP_GB}, correspond to two
rungs of the ribbon in Fig.~\ref{fig:Fbands}(a).

  Hence, the perturbed density of states and the 
contributions from  different basis functions can be 
calculated
\beq
\label{eq:PDOS}
\rho_{\mu m}(\epsilon) = -\frac{1}{\pi} \text{Im} 
\left[\sum_{\nu m'} G^{\mu\nu}_{m m'} 
(\epsilon) S_{\nu\mu,m'm} \right] \
\eeq 
as a decomposition of the total density of perturbed states 
$\rho(\epsilon)$.
  The latter, suitably normalised, does not differ from 
$\rho^0(\epsilon)$, 
given the infinitesimal weight of the scattering region. Hence,
  the decomposed functions are significant.
  Here we will use decomposition by site, showing
\beqs
\rho_{\mu}(\epsilon)= \sum_m \rho_{\mu m}(\epsilon) \, .
\eeqs

  Eq.~\eqref{eq:dyson_eq} is solved by matrix inversion 
numerically and converged results are obtained for a 
cutoff $m_c=50$, rendering
matrices of $202\times 202$, given the two sites 
directly affected by the projectile potential in 
Eq.~\eqref{Eq:VP_GB}.

  Figure \ref{fig:pert_dos_pot} shows the perturbed DOS
$\rho(\epsilon)$ projected on site $\mu=0$ 
in the presence of the projectile introduced in 
Eq.~\eqref{Eq:VP_GB} for a range of repulsive
(Fig.~\ref{fig:pert_dos_pot}.a-b) and attractive
(Fig.~\ref{fig:pert_dos_pot}.c-d) values of $\varepsilon_p$. 
  They show how the spectral weight of regions of 
large $\rho^0(\epsilon)$ (in the region between the
van Hove singularities) is shifted away, with peaks
appearing in the low-$\rho^0(\epsilon)$ region. 
 This is comparable to the localised state generated 
by a local perturbation in a static 1D TB chain
(see Fig.\ref{fig:TB} in Appendix~\ref{app:conventional_TB}).
For the moving system, however, a resonance appears instead 
of a localized state, given the fact that the unperturbed
spectrum has no gaps. 

  It is apparent in Fig.~\ref{fig:pert_dos_pot} that 
the spectral weight shift upwards for the repulsive 
projectile potential is different from the equivalent
shift downwards of the attractive counterpart,
breaking the up-down symmetry that appears in the
conventional locally perturbed TB (static) chain 
\cite{Economou}.
  As illustrated in Appendix~\ref{app:conventional_TB},
the static TB chain with a local perturbation also 
breaks that up-down symmetry whenever the perturbing
potential breaks inversion symmetry (left-right symmetry
in the chain): the usual picture of having $\rho_{\mu}
(-\epsilon)$ for $\varepsilon_P>0$ equal to 
$\rho_{\mu}(\epsilon)$ for $\varepsilon_P<0$,
and vice-versa (for $\epsilon=0$ in the middle of the TB 
band), does not hold when the perturbation is not
centrosymmetric either around an atom or around the center
of a bond.
 Therefore it is no surprise that we observe a similar
effect in the case of the moving projectile, since the motion
itself breaks that symmetry.
  
  The site-projected perturbed DOS is expected to recover 
that of  the unperturbed one when moving sufficiently far away 
from the projectile.
  This behaviour is shown in Fig.\ref{fig:pert_dos_site},
which displays $\rho_{\mu}(\epsilon)$ at various sites 
$\mu$ moving away from the projectile.

\subsection{Projectile velocity dependence}
\label{sec:projectile_velocity}

\begin{figure}[t]
\centering
\includegraphics[width=0.5\textwidth]{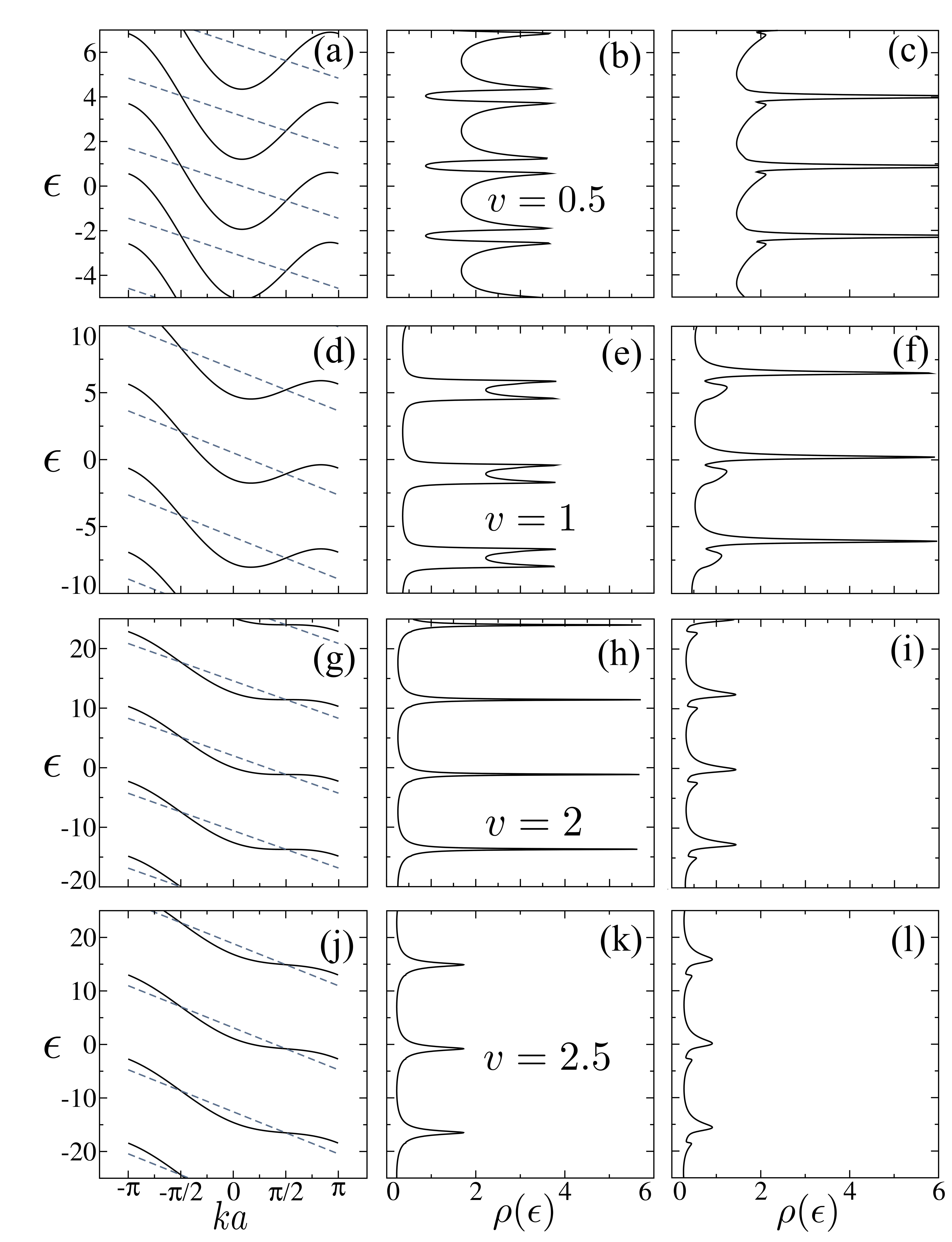}
\caption{Quasi-energy bands of scattering states, unperturbed 
DOS and  perturbed DOS for $v=0.5$ (a-c), $v=1.0$ (d-f), 
$v=2.0$ (g-i), and $v=2.5$ (j-l), all in units of 
  $\gamma a/\hbar$.
  $v=v_F=2\gamma a/\hbar$ is the highest group velocity of
electronic states in the LRF, and the Fermi velocity at 
half-filling. 
  $v_P.v_F$ define the supersonic regime.}
\label{fig:low_high}
\end{figure}

  Figure \ref{fig:low_high} shows the unperturbed and 
perturbed DOS of electronic stationary states for
$\varepsilon_P=0.7 \gamma$, for various
values of the projectile velocity, namely, $v=0.5$, 1.0, 
2.0 and $2.5$, in units of $v_0 \equiv \gamma a /\hbar$, 
along with the quasi-energy bands of the unperturbed crystal.
  As before, the periodically repeated bands have an 
energy separation of $\hbar \omega=2 \pi \hbar v/a$, 
proportional to the velocity of the projectile.

  For a velocity of $v=v_0$ (as in panels
d, e, and f of Fig.~\ref{fig:low_high}, and as in 
Fig.~\ref{fig:pert_dos_pot}), there are quasi-energies for 
which three asymptotic states are degenerate, allowing for 
e.g. an electron coming in from the right to be 
transmitted (same state), scattered back towards the left
or remain going to the right but more slowly (all in the PRF).
  For other values of the quasi-energy (the region with
lower $\rho^0(\epsilon)$) there is only one asymptotic
state and there is no scattering channel beyond
pure transmission: the projectile is transparent at
those quasi-energies.
    This is rather a peculiarity of the single band model,
since any more realistic model would include higher bands
which would provide scattering options for any 
quasi-energy and any projectile velocity.

  Increasing the velocity from our $v=v_0$ starting 
value, the cell grows, the van Hove singularities 
enclosing the three-state regime get closer to each other, 
until, for $v$ reaching the largest electronic group 
velocity (the Fermi velocity at half filling, 
$v_F = 2 v_0$), both van Hove singularities 
merge into one.
  Beyond that first critical velocity $v^c_1=v_F=2v_0$ 
the projectile is swifter than any electron, the supersonic regime, 
and no scattering process takes place for any quasi-energy,
again, a peculiarity of the single-band model.
  An example is shown in Fig.~\ref{fig:low_high}(j)-(l),
for $v_P=2.5\gamma a/\hbar$.
  The perturbed and unperturbed DOS locally differ, since
the projectile potential still affects the wave-functions
locally, but there is no outgoing Bloch wave different
from the incoming one regardless of which incoming one it is.

  For slower projectiles, the quasi-energy unit cell
becomes smaller, the van Hove singularities 
of the unperturbed DOS get closer together squeezing
the non-scattering region, thereby squeezing the resonance
in the perturbed DOS with them, as illustrated in
Fig.~\ref{fig:low_high} (a)-(c).
  If for $v=v_0$ there were regions of quasi-energy 
for which there were up to three compatible states,
slowing down below a critical velocity, $v^c_2$, an interval 
of quasi-energy values appear for which there are five degenerate
asymptotic states, below $v^c_3$ there are seven, and 
below $v^c_n$ there are $2n+1$, crowding towards 
the low-velocity limit, which becomes harder to treat, 
except for $v=0$ strictly, which becomes the much 
simpler static impurity problem.
It is a singular limit \cite{singular}, 
analogous to the one found when treating low-$k$ phonons
in a crystal, which becomes hard when addressing the 
periodic superlattice capturing their long wave-lengths, 
while it becomes trivial when strictly at the $\Gamma$ point.

  For velocity below $v^c_1=v_F$, the tilted bands of 
Fig.~\ref{fig:Fbands} (b) display local minima and maxima.
  Further critical values $v_n^c$ are defined by the velocities 
for which a horizontal line tangentially touches one minimum 
and one maximum (best seen as one single tilted band in 
an extended reciprocal space plot), which happens when
\bals
v_F \cos ka &= [(n+1/2)\pi -ka ] v \\
v_F \sin ka &= v \, .
\eals
  Solving for $k$ and $v$ yields
\beqs
v^c_n = v_F \left \{1, 0.219, 0.129, \dots , \, \sim \, 
\frac{1}{(n+1/2)\pi} \right \}
\eeqs
for $n\geq 1$ (the last expression being for large $n$).

\section{Particle density}

  The independent-particle problem discussed so far can 
then be used to address the many-particle problem
using a mean-field approach.
  The most attractive proposition given its efficiency 
and success in other contexts would be the
one based on Kohn-Sham (KS) time-dependent DFT 
\cite{runge,Marques2006}, 
  It has been shown, however, that Floquet TDDFT may be ill 
defined \cite{Maitra2002,Samal2006,MAITRA2007}.
  Nevertheless, the main experimental observable in the
field of electronic stopping processes is the electronic
stopping power, which relates to the suitable average of
the force opposing the
motion of the projectile, and which, as long as the
projectile potential is local, can be obtained
as the simple functional of the particle density
$n(\mathbf{r},t)$
\beqs
\mathbf{F}_P(t) = - \int \mathrm{d}^3\mathbf{r} \, \, 
n(\mathbf{r},t) \nabla V_P(\mathbf{r},t)
\eeqs
quite generally, regardless of the theory with which 
$n(\mathbf{r},t)$ is obtained (see e.g. the discussion
of Ehrenfest forces in \cite{Todorov2001}).
  We will just assume it is a mean-field theory, and
use the single-particle problem discussed in previous
sections to define the perturbed particle density.

\subsection{Occupation}
\label{sec:occupation}

  Two extra ingredients are needed beyond what obtained so far, 
occupation and self-consistency. 
  The latter is used to define the effective potential in the 
single-particle Hamiltonian iteratively
from the perturbed density and/or perturbed wave-functions. 
  However, in the context of this paper it only represents
a redefinition of the parameters defining the model.

  The occupation requires special attention.
  Occupation is normally quite trivially treated in equilibrium
or near equilibrium, 
by simply integrating the relevant Green's functions from 
$-\infty$ to the Fermi level.
  However, in our case, occupied states are defined 
by the Fermi level in the LRF, which means that 
occupation in the PRF is defined by a ``tilted Fermi level''
(shown in Fig.~\ref{fig:Fbands}). 
  Therefore, it depends on the crystal momentum $k$ of the 
unperturbed incoming scattering states.
  At any given quasi-energy there can be both occupied and 
unoccupied states, as seen in Fig.~\ref{fig:Fbands}.

\begin{figure}[t]
\centering
\includegraphics[width=0.40\textwidth]{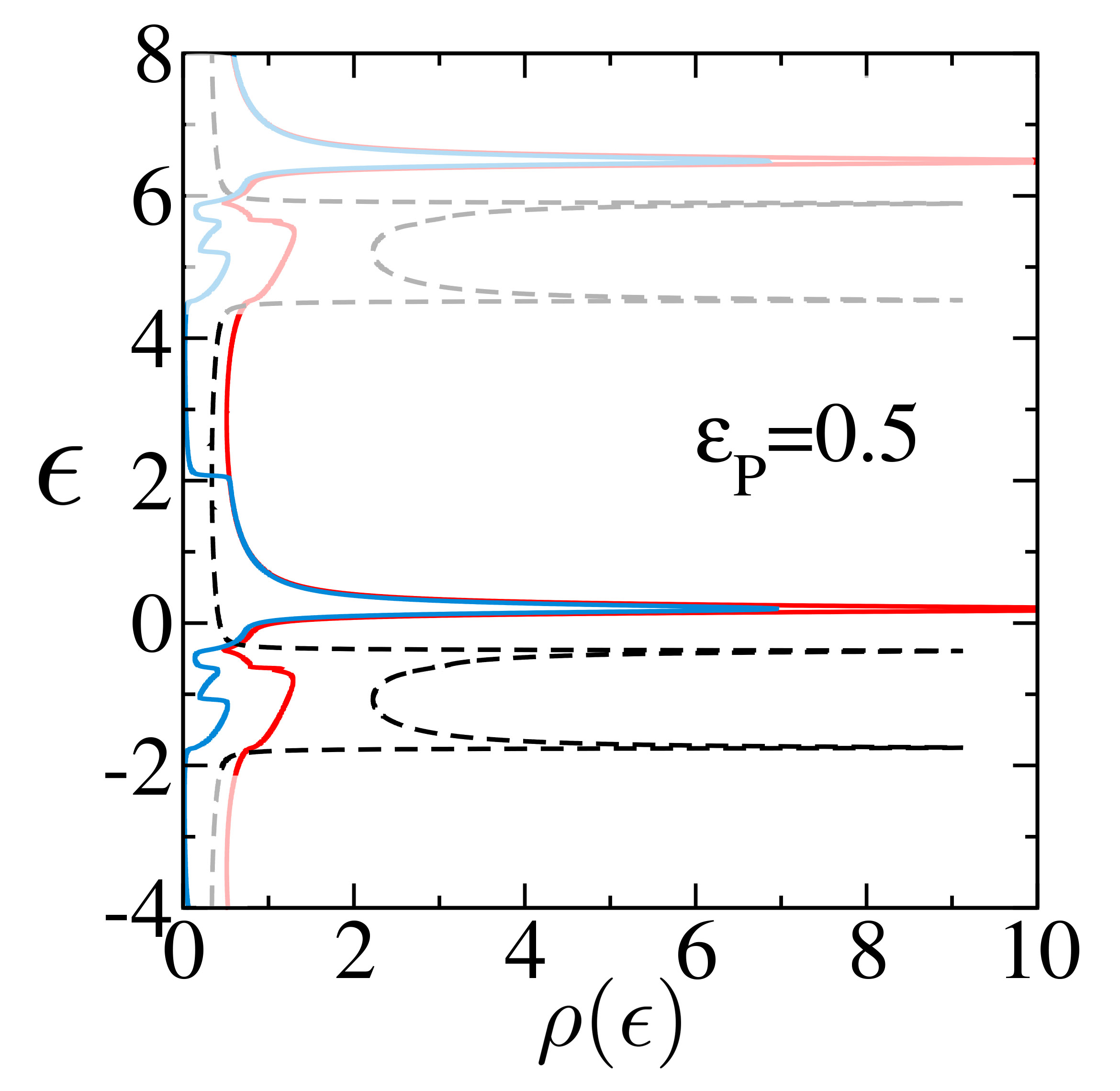}
\caption{Comparison between the unperturbed DOS (dashed black), 
perturbed DOS (red) and occupied perturbed DOS (blue), vs 
quasi-energy $\epsilon$ in units of $\gamma$,
for a projectile with velocity $v=v_0$ and perturbation 
potential $\varepsilon_p=0.5 \gamma$ (with 
$\eta= 0.01\gamma$).}
\label{fig:pert_dos_occupied}
\end{figure}

\begin{figure*}
\centering
\includegraphics[width=0.8\textwidth]{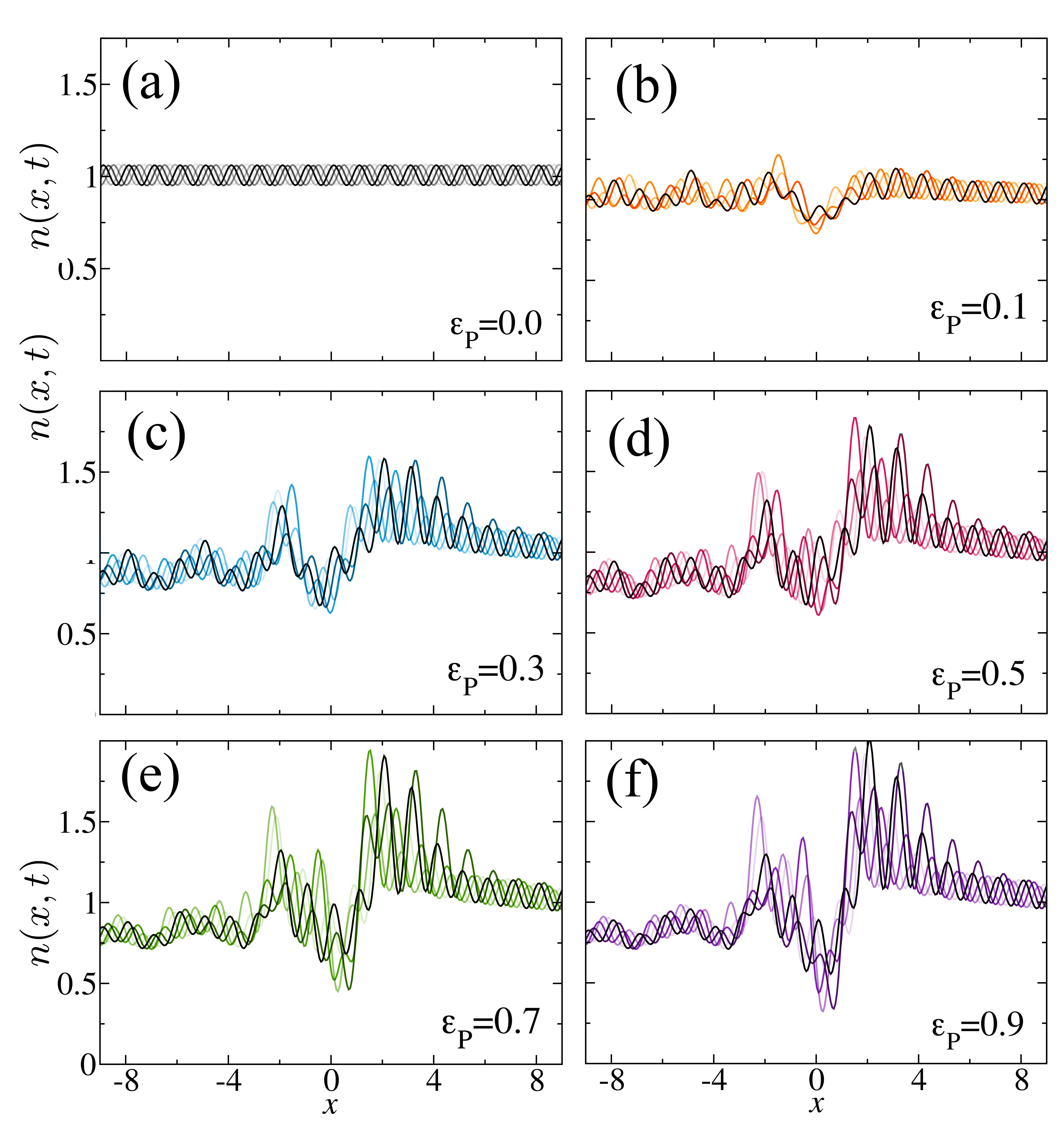}
\caption{Particle density $n(x,t)$ snapshots at $t = n \tau/5$
for $n=0,1,\dots 4$ (lighter curves for earlier times), for a
projectile of $\varepsilon_P=0,0.1\gamma,0.3\gamma,0.5\gamma,0.7\gamma$,
and $0.9\gamma$ in panels (a) to (f), respectively. 
  The velocity of the projectile is set to $v=v_0$}.
\label{fig:sigma_t}
\end{figure*}

  We address the  occupation problem analogously to 
earlier work for non-equilibrium ballistic
transport \cite{transiesta}, where equilibrium is defined 
separately in the two side electron reservoirs (leads),
thereby having two separate Fermi levels.
  The idea is to obtain the scattered wave-functions from 
the Lippmann-Schwinger equation, 
\beqs
|\Psi_{n,k}\rangle\rangle = \{\mathbb{1} + \mathbf{G}
[\varepsilon(k)] \mathbf{V}_P\} |\psi_{n,k}\rangle\rangle \, ,
\label{eq:pert_states}
\eeqs
for all incoming scattering states that correspond to occupied 
states in the laboratory frame.
We then use the scattered wave-functions to build the 
occupied Green's function (equivalent to the ``lesser''
Green's function in \cite{transiesta}), as
\beqs
\mathbf{G}^{<}(\epsilon) = \sum_{k}^{occ} \frac{|\Psi_k\rangle \rangle
\langle \langle \Psi_k |}{\epsilon - \varepsilon(k) + i \eta}
\eeqs
the sum running over all perturbed states $\Psi_{k}$
that result from the scattering of the initially occupied
asymptotic (Bloch) states of the crystal.
  Integrated many-particle quantities such as the 
particle density are then obtained by suitable integrals
of $\mathbf{G}^<(\varepsilon)$ over all quasi-energies.
  In our representation, 
\beqs  
G^{<\mu\nu}_{mm'}(\epsilon) = 
\sum_{k}^{occ} \frac{\Psi^{\mu}_{mk} \Psi^{\nu *}_{m'k}}
{\epsilon - \varepsilon(k) + i \eta} \, .
\eeqs 

  The density matrix, $D^{\mu\nu}_{mm'}$, 
defined as
\beq
\label{eq:Den_mat}
D^{\mu\nu}_{mm'}=\sum_k^{occ} \Psi^{\mu}_{mk} \Psi^{\nu*}_{m'k},
\eeq
can then be obtained from $\mathbf{G}^<$ by integrating
over all quasi-energies in one cell, 
\beqs  
D^{\mu\nu}_{m m'}= -\frac{1}{\pi} \text{Im} 
\int^{\hbar\omega} \! \! G^{<\mu\nu}_{mm'}
(\epsilon) \mathrm{d}\epsilon  \, ,
\eeqs
from which the particle density is obtained directly 
(Section~\ref{sec:density}),
as well as properties depending on it, such as forces on
atoms, and, from the force on the projectile, the
electronic stopping power.
  It is illustrative, however, to see the density of 
occupied states projected on the different sites, 
\beq
\label{eq:PDOSocc}
\rho_{\mu}^{occ}(\epsilon) = -\frac{1}{\pi} \text{Im} 
\left[\sum_{\nu m m'} G^{< \mu\nu}_{m m'} 
(\epsilon) S_{\nu\mu,m'm} \right] \, ,
\eeq 
which is equivalent to what displayed in 
Fig.~\ref{fig:pert_dos_pot},
but now for $\mathbf{G}^{<}$, and it shown in 
Fig.~\ref{fig:pert_dos_occupied}.

\subsection{Particle density $n(x,t)$}
\label{sec:density}

  The particle density $n(x,t)$ of the 1D chain in the 
presence of  the projectile in real space and time is 
given by
\beq
n(x,t)=\sum_{\mu\nu,mn}D^{\mu\nu}_{mn}
\xi_\mu(x,t) \xi_\nu^*(x,t) e^{i(m-n)\omega t},
\eeq
where the $\xi_\mu(x,t)$ are the gliding basis functions
as defined in Eq.~\eqref{eq:gliding}.

  The evolution of $n(x,t)$ during one time period ($\tau$) 
is shown in Fig.~\ref{fig:sigma_t} for various values of the 
perturbation potential. 
  The time evolution is indicated by superimposing snapshots
at $t=0, \tau/5, 2\tau/5, 3\tau/5$, and $4\tau/5$.
  Snapshots for subsequent times on the same sequence fall 
exactly on the depicted ones. 
  The implicit orthonormal basis functions of the original 
TB model have been given an explicit shape (see 
Appendix~\ref{app:basis-shape}) for the plotting of $n(x,t)$.

  The small wavelength oscillations depicted relate 
to the shape of orbitals, with the periodicity of the lattice, 
as can be seen in the absence of projectile, in panel (a) of
Fig.~\ref{fig:sigma_t}.
  As $\varepsilon_P$ is increased, a growing charge depletion 
is observed in the figure, around (and slightly in front of) 
the repulsive projectile at $x=0$.
  Since $v>0$, the projectile is moving to the right, and it 
is also apparent how the density is enhanced on the right of
the projectile and depleted on the left.
  The appearance of oscillations of larger wavelength 
than the lattice is also observed, in analogy with 
what happens in a static TB.
  A comparison with results for $v=0$ is provided
in Appendix~\ref{app:conventional_TB}.

%%%%%%%%%%%%%%       CONCLUSIONS       %%%%%%%%%%%%%%%

\section{Conclusions}
\label{conclusions}

  A local basis implementation of the Floquet theory
of electronic stopping of Ref.~\cite{Forcellini2019}
has been devised using a one-dimensional single-band
tight-binding model for demonstration, but also as a 
simple (simplest) model for describing the stroboscopically
stationary states resulting from electronic stopping processes
for projectiles of any strength and velocity.

  Once a gliding basis transformation is proposed to
define a time-periodic but not displacing basis set in 
the projectile reference frame, the single-particle 
scattering states are obtained with a conventional 
Dyson - Green's functions scattering formalism. 
  The integration over all incoming states for a 
determination of many-particle properties at a
mean-field level is accomplished by summing over
the perturbed scattering states from 
the occupied incoming ones using the Lippmann
Schwinger equation. 
  From the Green's function for occupied states 
the density matrix and the particle density are
readily obtained.
  
  Although both the jellium work \cite{echenique81,
Schonhammer1988,Bonig1989,Zaremba1995,Lifschitz1998} 
and its Floquet generalisation \cite{Forcellini2019} 
offer expressions for the electronic stopping power 
as key magnitude in comparison with experiment,
they are based on the individual single-particle scattering
amplitudes and corresponding single-particle energy changes
in the laboratory reference frame, which would be
perfectly adequate for a system of truly non-interacting
particles, but not for TDDFT (see e.g. 
Ref.~\cite{Nazarov2005}) or similar mean-field theories.
  The quasi-energy conserving individual Floquet scattering
states of the Kohn-Sham particles give a good
approximation to the particle density $n(x,t)$, 
however. 
  The stopping power can then be obtained at the same
level of theory directly from the force acting on the
projectile, which is an explicitly known functional of
the density \cite{Schmidt1996}, and which is 
straightforwardly calculated in any modern electronic
structure program.

%%%%%%%%%%%%%%       ACKNOWLEDGMENTS      %%%%%%%%%%%%%%%

\begin{acknowledgments}
  Funding from the Leverhulme Trust is acknowledged, 
under Research Project Grant No. RPG-2018-254, as well as 
from the EU through the ElectronStopping Grant Number 333813, 
within the Marie-Curie CIG program, and by the Research Executive 
Agency under the European Union's Horizon 2020 Research and 
Innovation programme (project ESC2RAD, grant agreement no. 776410).
  Funding from Spanish MINECO is also acknowledged, through 
grant FIS2015-64886-C5-1-P, and from Spanish MICIN
through grant PID2019-107338RB-C61 / AEI /DOI: 10.13039 / 501100011033.
  A UK's EPSRC studentship and and Grants No.
EP/L504920/1 and No. EP/N509620/1 are also acknowledged.
\end{acknowledgments}

%%%%%%%%%%%%%%       APPENDICES      %%%%%%%%%%%%%%%

\appendix  % Asterisk if only one appendix

\section{Moving tight-binding model in 3D}
\label{app:tb_general}

  The Bloch basis are constructed starting 
from the set of local basis $\phi'_{li}(\mathbf{r}') \equiv
\phi'_l(\mathbf{r}'-\mathbf{t}_i)$, where $l$ indicates
the orbital type and $\mathbf{t}_i$ is a vector indicating the 
center of the atom in the primitive unit cell (position $i$).
  The Bloch basis can be then defined as
\begin{equation}\label{eq:Bloch_basis}
\begin{split}
\chi'_{\mathbf{k}li}(\mathbf{r}') =& \frac{1}{\sqrt{N}}
\sum_{\mathbf{R}'}
e^{i\mathbf{k}\cdot\mathbf{R}'}\phi'_{li\mathbf{R}'}(\mathbf{r}').
\end{split}
\end{equation}
where the summation goes over all of the lattice vectors 
$\mathbf{R}' = \mu_1 \mathbf{a}_1+ \mu_2 \mathbf{a}_2+ 
\mu_3 \mathbf{a}_3$ as $\phi'_{li\mathbf{R}'}(\mathbf{r}')=
\phi_{l}(\mathbf{r}'-\mathbf{t}_i-\mathbf{R}')$.
  They can be used as the basis for the single-particle
eigenstates of the unperturbed crystal Hamiltonian
\begin{equation}\label{eq:Bloch_lab}
\lambda'_{n\mathbf{k}}(\mathbf{r}') = 
\sum_{l,i} c_{n\mathbf{k}li}\chi'_{\mathbf{k}li}(\mathbf{r}'),
\end{equation}
and are associated with eigenvalues $E_n(\mathbf{k})$ 
for band $n$.
  Once the crystal states are found, they can be transformed 
to the PRF via the Galilean transformation $\mathcal{G}$ 
as in the 1D case \cite{landau}.
\begin{equation}\label{eq:movingBloch3D}
\lambda_{n\mathbf{k}}(\mathbf{r},t) = \sum_{l,i} 
c_{n\mathbf{k}li} e^{-\frac{i}{\hbar}\left( E_n(\mathbf{k}) + 
\frac{1}{2}mv^2 \right)t}e^{-\frac{i}{\hbar}m\mathbf{v}\cdot 
\mathbf{r}}\chi'_{\mathbf{k}li}(\mathbf{r}+\mathbf{v}t) \, .
\end{equation}

  The Bloch-Floquet modes are
\beq \label{eq:BlochFloquetModes3D}
\begin{split}
&\psi_{n\mathbf{k}}(\mathbf{r},t) =
e^{-i\mathbf{k}\cdot\mathbf{v}t}e^{-
\frac{i}{\hbar}m\mathbf{v}\cdot\mathbf{r}}
\sum_{il}c_{n\mathbf{k}li} \ \chi'_{\mathbf{k}li}
(\mathbf{r}+\mathbf{v}t)\\
&= e^{-i\mathbf{k}\cdot\mathbf{v}t}
e^{-\frac{i}{\hbar}m\mathbf{v}\cdot\mathbf{r}}
\frac{1}{\sqrt{N}}\sum_{il\mathbf{R}}c_{n\mathbf{k}li} \ 
e^{i\mathbf{k}\cdot\mathbf{R}}\tilde{\phi}_{li\mathbf{R}}
(\mathbf{r},t),
\end{split}
\eeq
where Eq.~\eqref{eq:Bloch_basis} was used
and $\tilde{\phi}_{li\mathbf{R}}
(\mathbf{r},t) \equiv \phi'_{li\mathbf{R}}
(\mathbf{r}+\mathbf{v}t)$. 
Note that this expression is simply the generalization 
to 3D of the 1D Floquet modes of Eq.~\eqref{eq:BlochFloquetModes}.
 It is, in principle, valid for any direction of 
$\mathbf{v}=v\mathbf{\hat{v}}$, and the resulting Floquet modes are 
time-periodic with a period $\tau=a/v$, being $a$ the unit cell length of 
the crystal repetition along the projectile's trajectory, which depends on 
the relative disposition of the trajectory and the host's crystal structure. 
  $a$ can therefore take values from the length of the shortest lattice 
vector, all the way to infinity.
  The latter case will arise along incommensurate directions in
the crystal, in which case the boosted Bloch states of 
Eq.~\eqref{eq:BlochFloquetModes3D} are not strictly Floquet modes 
since they are not time periodic ($\tau\rightarrow\infty$).

  The Floquet modes of Eq. \eqref{eq:BlochFloquetModes3D} are 
eigenstates of the Floquet operator $\mathcal{H}(\mathbf{r},t) = 
H(\mathbf{r},t) -i\hbar\partial_t$.
  Similarly, the localised gliding basis set (per orbital type) can
be defined in the direction of the velocity 
\begin{equation}
\begin{split}
\xi_{li\mathbf{R}}(\mathbf{r},t) = {\cal N}(t)&[f(\delta t) 
\phi_{li\mathbf{R}+na\mathbf{\hat{v}}} (x,t) \\
&+ f(\delta t - \tau)\phi_{li\mathbf{R}+(n+1)a\mathbf{\hat{v}}} (x,t)],
\end{split}
\end{equation}
where all the definitions from Sec.~\ref{sec:gliding}
carry on unchanged.
It should be noted that in 2D and 3D,  other definitions for the gliding basis might 
be more convenient when considering velocity directions deviating from the primitive 
lattice vectors.
  In addition, depending on the exact direction of $\mathbf{v}$,
the treatment can become numerically very complex --see the 
discussion in \cite{ForcelliniThesis}, p. 69, analogous to
the $v\rightarrow 0$ limit (Section~\ref{sec:projectile_velocity})-- 
and other methods (or particularly tailored gauge choices) could 
be more suitable.

\section{Floquet modes from the Bloch states}  
\label{app:movsol}

  Consider the moving Bloch state in the PRF
\beqas
\lambda_k(x,t) = e^{-i \frac{mv}{\hbar}x} %& 
e^{-i [E(k) + \frac{1}{2} mv^2] t/\hbar} %\\
%& 
\sum_{\mu} \frac{e^{ika\mu}}{\sqrt{N}} \tilde{\phi}_{\mu}(x,t) \, .
\eeqas
  In the Bloch form, from Ref. \cite{Forcellini2019}
\beqs
\lambda_k(x,t) = e^{-i [E(k) + \frac{1}{2} mv^2 - \hbar kv ] t/\hbar} 
\left [ e^{ix(k-mv/\hbar)} u_k(x,t) \right ] \, ,
\eeqs
where $u_k(x,t) \equiv u'_k(x+vt)$ is
the periodic envelope of the Bloch eigenstate,
thus defining the time-periodic  mode $\psi_k(x,t)$ as the
expression in brackets.
  By comparing the two expressions the Bloch-Floquet mode in this 
local basis representation is readily extracted
\beqas \label{eq:blochfloquet1}
\psi_k(x,t) = e^{-ikvt}e^{-imvx/\hbar}\frac{1}{\sqrt{N}}\sum_{\mu} 
e^{i\mu ka} \tilde{\phi}_{\mu}(x,t),
\eeqas
periodic with the period of $\tau = a/v$ and
eigenstate of the Floquet Hamiltonian 
$\mathcal{H}(x,t) = H(x,t) -i\hbar{\partial_t}$.
  Indeed, performing the direct calculation for $\psi_k(x,t)$ 
[using the simplified notation $f \equiv f(x,t)$]
\beqs
-i\hbar\frac{\partial}{\partial t}\psi_k = -\hbar kv\psi_k 
+ v e^{-i(kvt +mvx/\hbar)} \hat{p}\frac{1}{\sqrt{N}}\sum_{\mu}
e^{i\mu ka} \tilde{\phi}_{\mu},
\eeqs
with $\hat{p} = -i\hbar {\partial_x}$.
By applying the Hamiltonian transformed in the PRF
$H(x,t) = H'(x,t) -v\hat{p}-mv^2/2$ to $\psi_k$
[where $H'(x')$ is the Hamiltonian in the LRF,
    $x=x'-vt$ using our convention]
one obtains
\begin{equation*}\label{eq:hamtransf}
\begin{split}
H\psi_k = &\left[E(k) + m v^2/2\right]\psi_k -v 
e^{-i(kvt +mvx/\hbar)} \\
&\times
\hat{p} \frac{1}{\sqrt{N}}\sum_{\mu}e^{i\mu ka}\tilde{\phi}_{\mu},
\end{split}
\end{equation*}
with $E(k)$ being the energy
of the Bloch state in the LRF.
The above lead directly to the Floquet equation
\beqas\label{eq:floquet_ev}
\mathcal{H}(x,t) \psi_k(x,t) = \left[E(k) + \frac{1}{2} mv^2 - 
\hbar kv \right]\psi_k(x,t),
\eeqas
verifying explicitly that $\psi_k$ is indeed the Floquet 
mode with the correct value for the quasi-energy.
The quasi-energy replicas arise from the 
$k$ values in the extended zone, since
\beqs
k \rightarrow k+n\frac{2\pi}{a} \; \Rightarrow \;
\varepsilon(k) \rightarrow \varepsilon(k) - 
\hbar \left ( n \frac{2\pi}{a} \right ) v 
= \varepsilon(k) - n \hbar \omega \, .
\eeqs

\begin{figure}[t] % it can be [tbxH!] 
\includegraphics[width=0.4\textwidth]{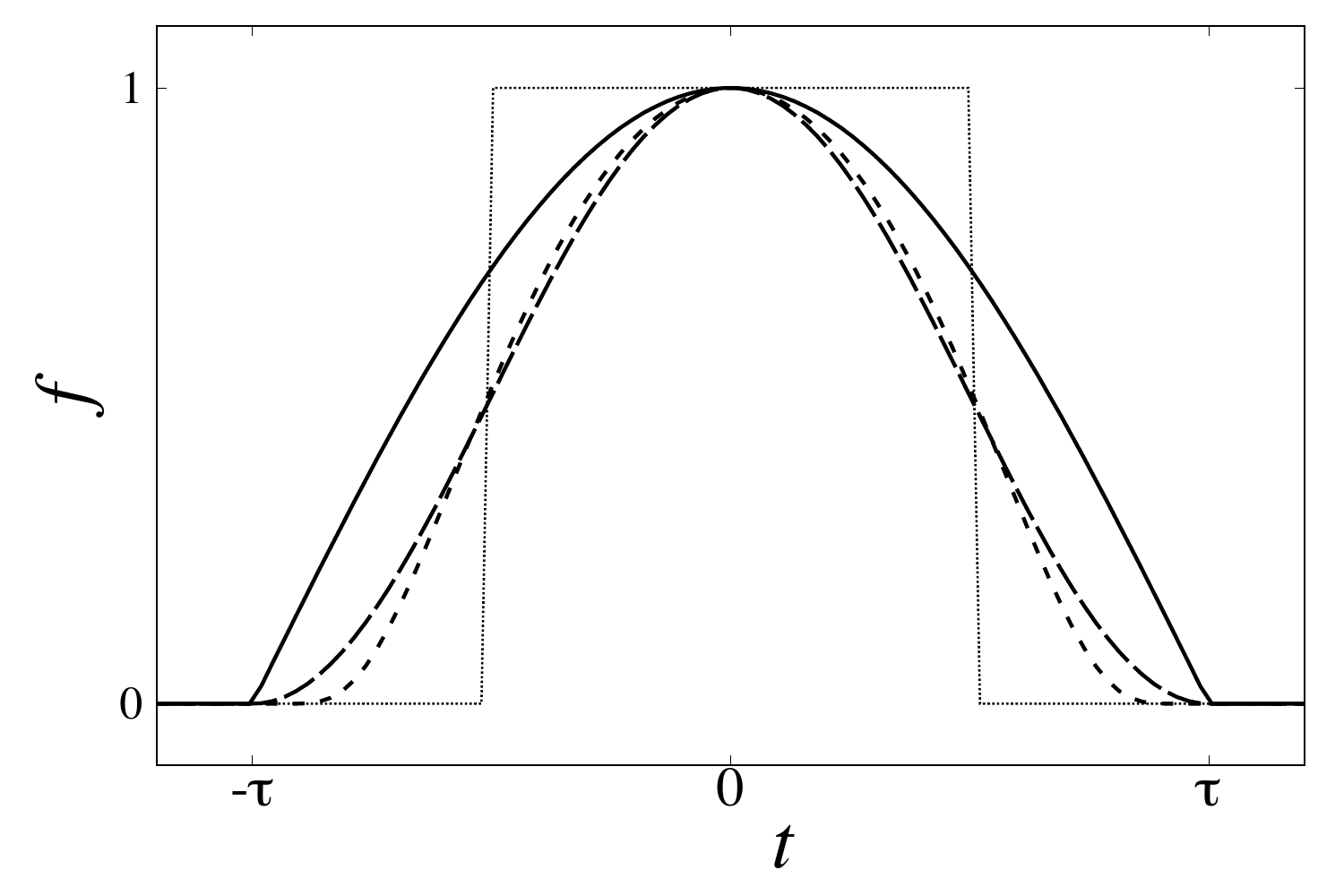}
\caption{Examples of gauge function $f(t)$, corresponding to the
four in Eq.~\ref{eq:gauge}, indicated by increasingly short dashes. 
$f_3(t)$ is depicted for $\alpha=2$.}
\label{fig:gauge}
\end{figure}

\section{Gauge functions for the gliding basis}
\label{app:fs}
   
  The function $f(t)$ defining the gliding basis transformation
represents a gauge freedom that can be used for convenience.
  Here a few examples:  
\begin{align}
\label{eq:gauge}
& \tilde{f}_1(t) = \left| \cos \left ( \frac{\pi}{2}\frac{t}{\tau} 
  \right ) \right|\nonumber \\
& \tilde{f}_2(t) = \cos^2 \left ( \frac{\pi}{2}\frac{t}{\tau} \right ) \\
& \tilde{f}_3(t) = e^{-\alpha t^2/(\tau^2-t^2)}  \nonumber \\
& \tilde{f}_4(t) = \Theta(t\!+\!\tau/2) - \Theta(t\!-\!\tau/2) \; , 
\nonumber 
\end{align}
where $\Theta(t)$ is the Heaviside step function,
and where $f(t)$ is defined from $\tilde{f}(t)$ as
\beqs
f(t) = \left \{ 
\begin{array}{ll}
 \tilde{f}(t)  & \; ,  \;  t \in [-\tau,\tau) \\
 0  & \; , \; t \notin [-\tau,\tau)
\end{array} 
\right .
\eeqs
  The the corresponding $f(t)$ functions are depicted in 
Fig.~\ref{fig:gauge}.
  $f_1(t)$ is convenient for simplicity, since ${\cal N}(t)=1$
at all times, but shows a derivative discontinuity at $t=\pm\tau$, 
while $f_2(t)$ displays continuity of the function and first 
derivative, with a discontinuous curvature at $t=\pm\tau$.
  $f_3(t)$ has all derivatives continuous there, and contains 
the free $\alpha$ parameter that fattens the function
within its limits.
  $f_4(t)$ gives the simplest, ``relabelling" transformation, i.e.,
$\xi$ orbitals follow the $\phi$ orbitals leftwise, but every period 
they abruptly jump by one lattice parameter right-wise. 
  That is, $\xi_{\mu}(x,t) = \phi_{\mu}(x,\delta t\!\!-\!\!\tau/2)$.
 $f_2(t)$ is the one used in the calculations presented 
in this work.

\section{Alternative projectiles}
\label{app:wrong-projectile}

  If the projectile operator $V_P$ is defined directly on the 
gliding basis (Section~\ref{sec:glide-projectile}), as, e.g.
\beqs
\langle \xi_{\mu} | V_P | \xi_{\nu}\rangle = \varepsilon_p 
\delta_{\mu\nu} \delta_{\mu 0} \, ,
\eeqs
knowing its form in the original basis requires the determination 
of the inverse of the basis transformation of Eq.~\eqref{eq:gliding}.

\subsection{Inverse transformation tensor}  
\label{app:transformation}

  The basis set transformation in Eq.~\eqref{eq:gliding}
can be expressed as
\beq
\label{eq:glitrans}
|\xi_{\mu} \rangle = |e_{\sigma}\rangle A^{\sigma}_{\pe\mu} 
\eeq
assuming summation over repeated indices and dropping the time 
dependence for brevity.
\beqs
A^{\sigma}_{\pe\mu} =  \langle e^{\sigma} |\xi_{\mu} \rangle
\eeqs
and $\{|e^{\sigma}\rangle, \forall \sigma=1\dots {\cal N}\}$ is 
the dual basis of $\{|e_{\sigma}\rangle\}$, such that 
$\langle e^{\sigma} | e_{\delta}\rangle = \langle e_{\delta} | 
e^{\sigma}\rangle =\delta^{\sigma}_{\delta}$,
and, consequently, $| e^{\sigma} \rangle \langle e_{\sigma}| =
| e_{\sigma} \rangle \langle e^{\sigma}| = P_{\Omega}$,
the projector onto the subspace spanned by the basis.
  In this case 
\beqs
A^{\sigma}_{\pe\mu}(t) = {\cal N}(t) [ f(\delta t) 
\delta^{\sigma}_{\pe\mu + n} 
+  f(\delta t-\tau) \delta^{\sigma}_{\pe\mu +n+1}]
\eeqs
which is a square matrix with a non-zero bi-diagonal that 
displaces leftwards and downwards.
  For simplicity in the following, let us re-express it, 
for any given time, as
\beq
\label{eq:short-glitransf}
A^{\sigma}_{\pe\mu} = c \, \delta^{\sigma}_{\pe\mu + n} 
+  s \, \delta^{\sigma}_{\pe\mu +n+1} \, .
\eeq

  The inverse transform is defined as 
\beq
\label{eq:glitrans2}
|e_{\sigma}\rangle = | \xi_{\mu} \rangle B^{\mu}_{\pe\sigma}
\eeq
with $B^{\mu}_{\pe\sigma}=\langle \xi^{\mu} | e_{\sigma} \rangle$.
  If both bases were orthonormal, the transformation matrix would be 
unitary, so that $B=A^{-1}=A^+$. 
  Since the gliding basis is not orthogonal, however, the 
inverse relations are $BA=AB=1$ in the sense
\beq
\label{eq:inverse}
A^{\sigma}_{\pe\mu} B^{\mu}_{\pe\gamma} = \delta^{\sigma}_{\gamma} 
\quad \mathrm{ and } \quad B^{\mu}_{\pe\sigma} A^{\sigma}_{\pe\nu} 
= \delta^{\mu}_{\nu} \; .
\eeq

  The sought expression for the original-basis representation of $V_P$ 
depends on the inverse transformation, since we need to express 
any $|e_{\sigma}\rangle$ in terms of the $|\xi_{\mu}\rangle$ states, 
and that is precisely Eq.~\eqref{eq:glitrans2}.
  Putting together Eqs.~\eqref{eq:short-glitransf} and 
\eqref{eq:inverse}, we obtain 
\beqs
c\, B^{\mu}_{\pe\sigma} + s B^{\mu}_{\pe\sigma-1} = 
\delta^{\mu}_{\pe\sigma} \, ,
\eeqs
and, again, for clarity, let us focus on $\mu=0$ and call
$B^{0}_{\pe\sigma}$ as $B_n$, giving the recursive relation
\beqs
c\, B_n + s B_{n-1} = \delta^{0}_{n} \, .
\eeqs
   For $c>s$ the solution is
\beqs
B_n = \left \{ \begin{array}{cc}
0 & n<0 \\ \\ \frac{1}{c} & n=0 \\ \\ \frac{1}{c} \left ( 
-\frac{s}{c} \right )^n & n>0
\end{array} \right . 
\eeqs
whereas for $c<s$,
\beqs
B_n = \left \{ \begin{array}{cc}
\frac{1}{s} \left ( -\frac{c}{s}\right )^n & n<-1 \\ \\ 
\frac{1}{s} & n=-1 \\ \\  0  & n>-1
\end{array} \right .
\eeqs
  That is, the lower (upper) triangle of the infinite matrix 
is zero for $c>s$ ($c<s$), while the elements of the other 
triangle display a sign alternation when moving away from the 
diagonal, with an exponential decay of the magnitude, 
\beqs
|B_n| \propto e^{-\zeta n}  \, ,
\eeqs
with $\zeta = \log (c/s)$ for $c>s$ and $\zeta = \log (s/c)$ 
for $c<s$.
  The decay length diverges when $c$ approaches $s$, swapping 
triangle precisely at $c=s$.
  Since $c$ and $s$ represent a periodic function in time, 
one delayed with respect to the other, the $B$ tensor starts diagonal, 
gradually extends into the upper triangle until full, then abruptly 
swaps into the full lower, which then gradually shrinks towards 
diagonal again (but shifted by one). 
  And so it cycles.

\subsection{Resulting projectile representation}

  The projectile potential expressed as $V_P=|\xi^0\rangle 
\varepsilon_p \langle \xi^0|$, becomes
\beqs
V_{P,\sigma\lambda} =\langle e_{\sigma}|V_P| e_{\lambda} \rangle
= \varepsilon_p \, B^{0\pe *}_{\pe\sigma} B^{0}_{\pe\lambda}
\eeqs
which gives a matrix with an exponential decay towards 
the lower-right quadrant and zero otherwise, the range then
diverging as $t$ approaches $n\tau$, and then
swapping to the opposite upper-left quadrant. 

  This behaviour is nonphysical and produces awkward behaviours.
  It is a rather unfortunate and non-intuitive 
consequence of establishing local decompositions in the 
non-orthogonal gliding transformation.
  But remember that the alternative discontinuous-relabelling 
basis transformation 
\beqs
|\xi_{\mu}, t \rangle = |e_{\mu}, \delta t\rangle
\eeqs
is nothing by a particular choice of gauge function in the 
gliding transformation (a step function), and that the situation 
at mid-period is abrupt filling and swapping as well, although it
may appear less explicitly.
  
  The situation for the natural representation choice
$V_P=|\xi_0\rangle \varepsilon_p \langle \xi^0|$ is less
symmetric but ultimately suffering from the same
oscillations in projectile-potential spatial range.
  This is why we have chosen to use the projectile as expressed
in Eqs.~\eqref{eq:projectile-LRF} and \eqref{Eq:VP_GB}.

\begin{figure}
\includegraphics[trim=0cm 0.cm 0cm 0cm,
clip=true,width=0.5\textwidth]{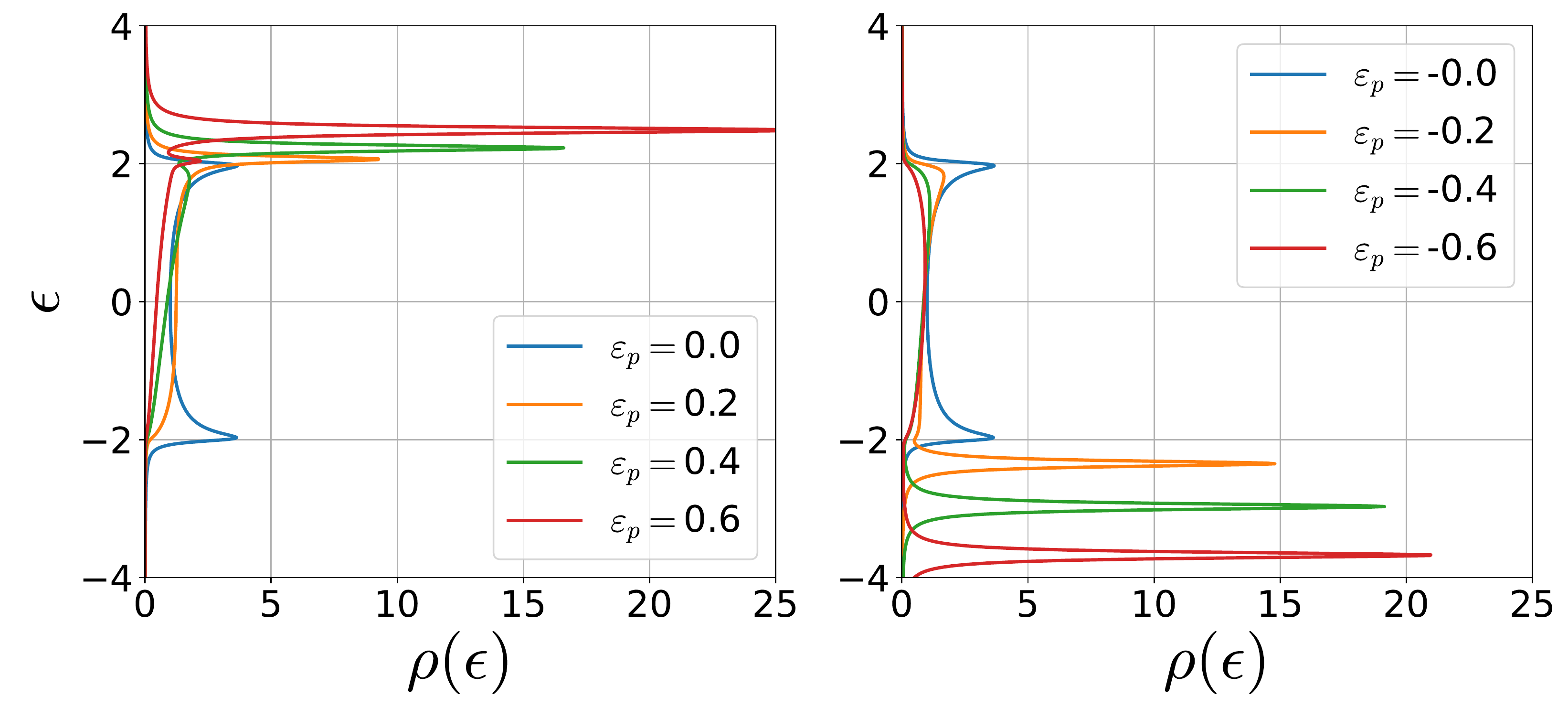}
\caption{Perturbed density of states $\rho$ on site 0 
vs energy $\epsilon$, in units of $\gamma$, 
for a static single-band 1D tight-binding model with a 
perturbation potential of the form specified in 
Eq.~\eqref{eq:asympert}, for various values of 
$\varepsilon_P$ (in units of $\gamma$) between 
$0.0$ (black) and $0.6$ (purple) on the left panel, 
and between $0.0$ (black) to  $-0.6$ (purple) on the right
panel.}
\label{fig:TB}
\end{figure}

\section{Static tight-binding impurity problem}
\label{app:conventional_TB}

\begin{figure}
\includegraphics[trim=0cm 0.cm 0.0cm 0.cm, 
clip=true,width=0.47\textwidth]{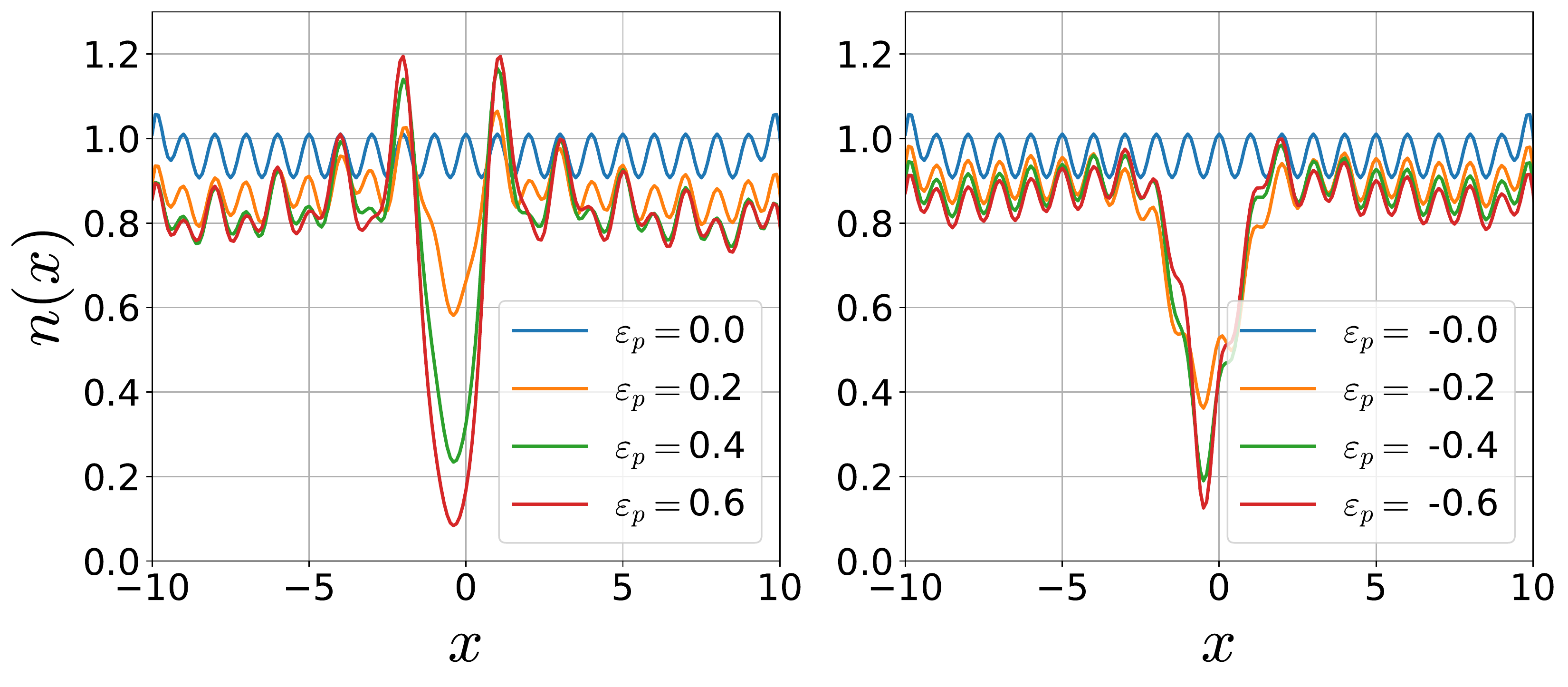}
\caption{Particle density $n(x)$ in real space for the 
same model and same values as Fig.~\ref{fig:TB}.
It has been obtained following the
procedure using the Lippmann-Schwinger equation
described in Section~\ref{sec:occupation}, and 
therefore, it does not include the particle density
associated to the bound state (note relevant only to the 
static attractive case).}
\label{fig:TB_charge}
\end{figure}

\begin{figure}[b]
\includegraphics[trim=0cm 0.cm 0.0cm 0.cm, 
clip=true,width=0.47\textwidth]{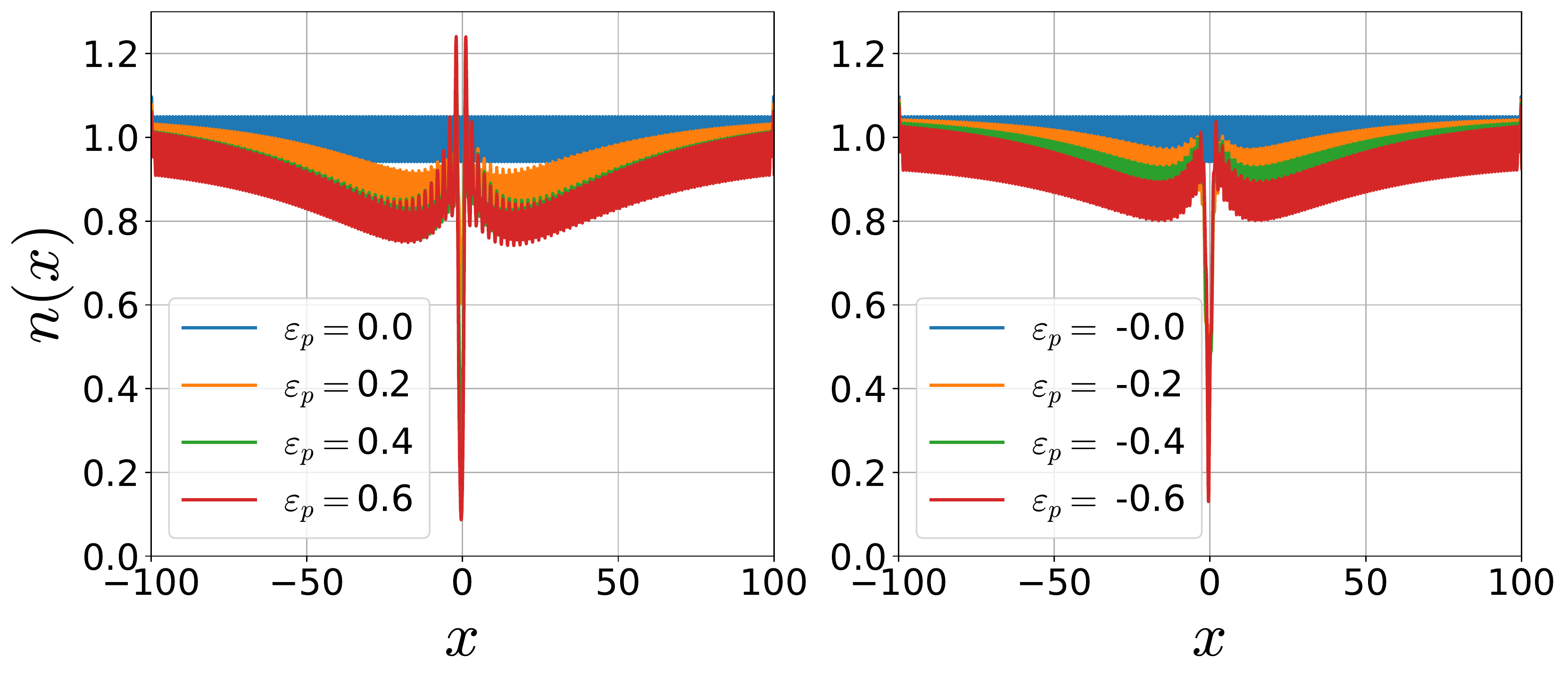}
\caption{Same as Fig.~\ref{fig:TB_charge} over a longer 
range in real space.}
\label{fig:TB_charge_l}
\end{figure}

\begin{figure*}
\centering
\includegraphics[width=0.8\textwidth]{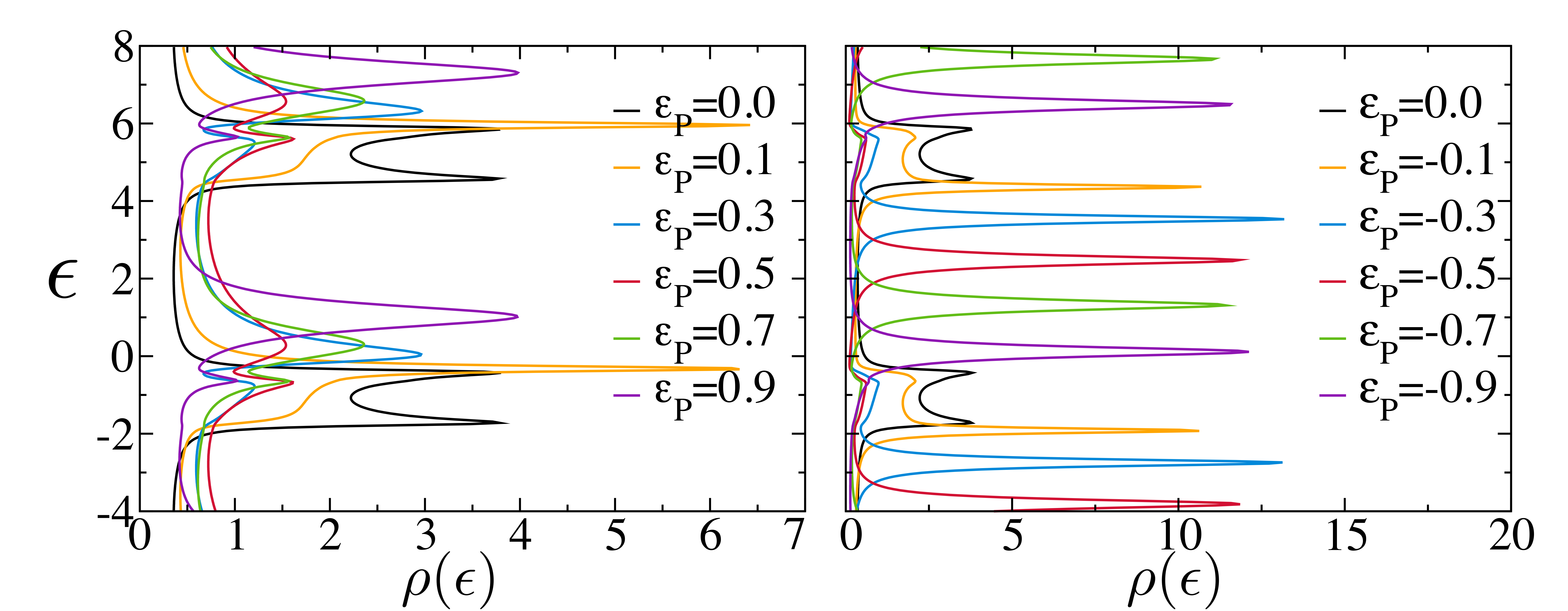}
\caption{Perturbed density of states $\rho$ vs quasi-energy
$\epsilon$ for $v=v_0$, calculated using the 
alternative definition of the projectile potential as 
shown in Eq.~\eqref{eq:3projectile}, which is defined directly 
on the gliding basis, using the gauge specified as $f_2$ in
Eq.~\eqref{eq:gauge} in Appendix~\ref{app:fs}.
Left (right) panel shows DOS for several positive (negative)
values of the perturbation potential strength $\varepsilon_P$.}
\label{fig:model2_perturbed_DOS}
\end{figure*}

  For a better insight into features of the perturbed 
density of states of the Floquet scattering model, especially 
related to symmetry, the results of similar calculations for
$v=0$ are presented here, i.e., a local impurity in a 1D 
single-band static tight-biding model.
  A local on-site impurity perturbation of the form
\beqs
V=|\phi_0\rangle \varepsilon_P \langle \phi_0 |
\eeqs
(LRF and PRF coincide for $v=0$) is known to produce 
a bound state above (below) the band for positive (negative)
$\varepsilon_P$, and a local density of states on the
perturbed site $\rho_0(\epsilon)$ which becomes 
$\rho_0(-\epsilon)$ when changing the sign of $\varepsilon_P$,
an ``up-down'' symmetry that is not observed in 
Fig.~\ref{fig:pert_dos_pot} of the Floquet model.
  
  That up-down symmetry in the static model is very
characteristic and related to the simplicity of the
model, with a very exceptional up-down symmetry in the
unperturbed density of states, plus the inversion symmetry
in space implied by the defined impurity potential $V$.
  Indeed, that symmetry is still observed when introducing
an off-diagonal $V$ instead
\beqs
V=(|\phi_0\rangle \gamma_P \langle \phi_1 | +\mathrm{h. c.})
\eeqs  
(h.c. standing for Hermitian conjugate), or a combination
of diagonal and non-diagonal (presenting the non-zero $V$ 
matrix block, for sites 0 and 1)
\beqs
 V=\begin{pmatrix} 
 \varepsilon_P & \gamma_P \\ \gamma_P^* & \varepsilon_P
  \end{pmatrix} \, ,
\eeqs
now preserving inversion symmetry around the center of
the bond between sites 0 and 1.

  However, the up-down symmetry disappears when
the inversion symmetry is broken, which is simplest to
describe with 
\beq
\label{eq:asympert}
V=\begin{pmatrix} 
\varepsilon_{P0} & \gamma_P \\ \gamma_P^* & \varepsilon_{P1}
\end{pmatrix} 
= \varepsilon_P \begin{pmatrix}
0.5 & 0.2 \\ 0.2 & 0.35
\end{pmatrix} 
\, ,
\eeq
where $\varepsilon_{P0}\neq \varepsilon_{P1}$, and where
a set of particular values are proposed 
scaled by a single parameter $\varepsilon_P$.
  Fig.~\ref{fig:TB} shows the perturbed density of 
states at site 0 for the specified perturbation,
for various values of the impurity potential 
strength $\varepsilon_P$.
  The up-down symmetry is visibly broken.

  The analogous difference for attractive versus repulsive 
local perturbation for the Floquet model apparent in 
Fig.~\ref{fig:pert_dos_pot} relates to the same inversion 
symmetry breaking, although in the Floquet case it is due to 
the right to left motion of the crystal with respect to the 
projectile in the PRF.

  The effect of the perturbing potential of 
Eq.~\eqref{eq:asympert} on the particle density of the static 
chain is presented in Fig.\ref{fig:TB_charge} for a range of 
values of $\varepsilon_P$, repulsive on the left panel and 
attractive on the right. 
 The functional form of the basis functions used for
that plot is defined in Appendix~\ref{app:basis-shape}.
 The same particle density is shown over a longer 
range in real space in Fig.\ref{fig:TB_charge_l}, 
showing the long-range perturbation characteristic
in 1D.

\section{Alternative projectile definition}

  Figure~\ref{fig:model2_perturbed_DOS} shows the same 
information as Fig.~\ref{fig:pert_dos_pot} but for the
alternative definition of the projectile perturbing
potential proposed in Eq.~\eqref{eq:3projectile}.
  It is defined directly on the gliding basis, which means
that it is gauge-dependent, and, although convenient to
write down, quite inconveniently dependent on the
arbitrary choice of gauge, which is $f_2$ of 
Eq.~\eqref{eq:gauge} in this case.
  The qualitative behavior is however unchanged.

\section{Local basis in real space}
\label{app:basis-shape}

  For the purpose of calculating the charge density 
in real space for producing Fig.~\ref{fig:sigma_t}, 
the real-space shape of the function $\phi(x)$ that gives
rise to the original basis set of the tight-binding model
$\{\phi_{\mu}, \mu \in Z\}$ has to be specified.
  It is defined as
\beqs
\phi(x) = \mathcal{N} e^{-\alpha x^2} \cos( \frac{2\pi}{b} x)
\eeqs
where $\alpha$ defines the width of the Gaussian, and $\frac{2\pi}{b}$
originates an underlying oscillation that ensures (and $b$ is chosen
such) that the nearest neighbor overlap is zero. 
  For $\alpha \lesssim a$ the second nearest neighbor overlap 
is not zero but negligible, giving an effectively orthonormal 
basis, once $\phi(x)$ is suitably normalized with
\beqs
\mathcal{N} = \left ( \frac{8\alpha}{\pi} \right )^{1/4}
(1 + e^{-2\pi^2/\alpha b^2})^{-1/2}.
\eeqs
  The values used in this work for Figs.~\ref{fig:sigma_t},
\ref{fig:TB_charge}, and \ref{fig:TB_charge_l} are $\alpha = a$ 
and $b=4 a$.

%%%%%%%%%%%%%%       Scattering theory      %%%%%%%%%%%%%%%

\section{Scattering Amplitudes}
\label{app:scattering}

\begin{figure*}
\centering
\includegraphics[width=0.7\textwidth]{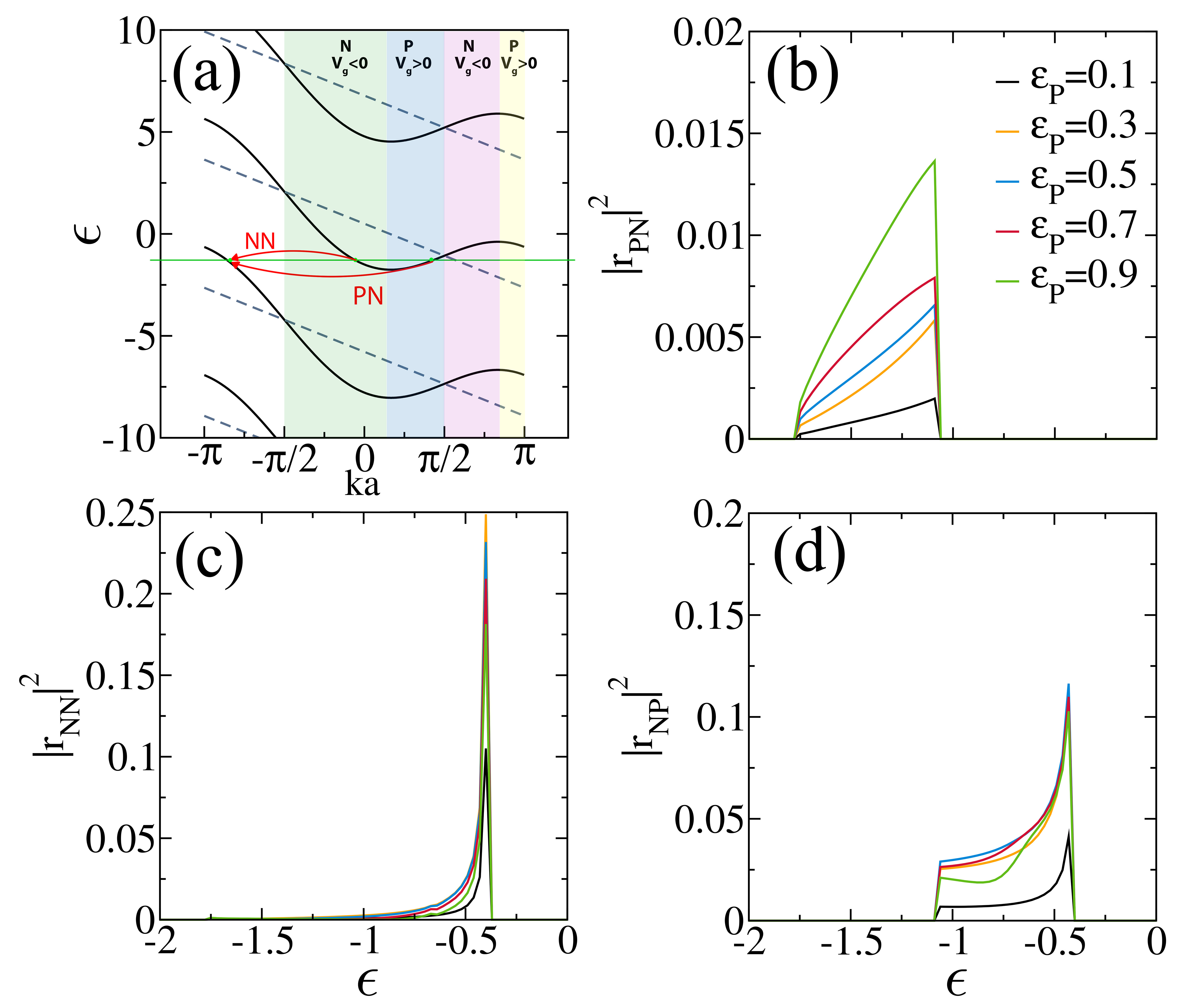}
\caption{Scattering matrix for $v=v_0$.
(a) Schematic illustration for the labeling of the 
scattering probability coefficients, $|r_{ij}|^2$.
  Scattering coefficients: (b) $|r_{PN}|^2$ from 
states with $v_g>0$ (right going states in the PRF, positive, P)
to states with $v_g<0$ (negative, N), (c) $|r_{NN}|^2$
with $v_g<0$ to states with $v_g<0$ and (d) $|r_{NP}|^2$ with 
$v_g<0$ to states with $v_g>0$, the three scattering possibilities
beyond transmission for $v=v_0$, scattering from occupied states.}
\label{fig:scat_amp}
\end{figure*}

  As argued in the paper, the safest way to obtain the
electronic stopping power and characterize the 
electronic distortion is via the particle density,
which is addressed in Section~\ref{sec:density}.
  However, the single particle description can also
render useful information for further analysis,
such as excitation rates for the different 
single-particle excitation channels involved, as 
well as the traditional estimation of the stopping power
directly from elementary processes and their
single-particle energy jump in LRF, as used in
Refs.~\cite{echenique81} and \cite{Forcellini2019}.
  The calculation of single-particle scattering amplitudes 
are presented here, from which such single-particle
results can be extracted.

Starting from the transition operator or $\mathbf{T}$-matrix
for our model 
\cite{Marinov_1996}
\beq
\label{eq:Tmat}
\mathbf{T}(\epsilon) = [ \mathbf{V}_P+ \mathbf{V}_P\mathbf{G}(\epsilon) 
\mathbf{V}_P ] \, ,
\eeq
the scattering matrix is then defined as 
\beq
\label{eq:Smat}
\mathbf{S}=\mathbb{1}- i \frac{a}{\hbar v_g}\mathbf{T} \, .
\eeq

  Scattering amplitudes have been calculated for $v=v_0$, 
for scattering from states with negative and positive (’N’ and 'P') 
group velocities to states with the same quasi-energy of 
negative and positive (’N’ and 'P') group velocities, 
which are separately shown in Fig.~\ref{fig:scat_amp}.
  A visual aid for the labeling of the scattering amplitudes 
is  depicted in Fig.~\ref{fig:scat_amp} (a).
  The magnitude of the scattering generally increases as the 
strength of the perturbation potential grows, as expected,
although the situation is considerably richer than the customary
1D reflection and transmission coefficients, including the quite
counter-intuitive behavior of perfect transparency for some 
incoming quasi-energies regardless of the strength of the 
projectile perturbation.

{54}%

\end{document}